\documentclass[aps,prd,10pt,nofootinbib,twocolumn,eqsecnum,showpacs,showkeys,superscriptaddress,preprintnumbers,altaffilletter]{revtex4-2}
\usepackage{graphicx}
\usepackage{amsmath}
\usepackage{multirow}
\usepackage{hyperref}
\usepackage{soul}
\usepackage{graphicx}
\usepackage{dcolumn}
\usepackage{amssymb}
\usepackage{amsfonts}
\usepackage{amsbsy}
\usepackage{color}
\usepackage{rotating}
\usepackage[english]{babel}
\usepackage{multirow}
\usepackage{float}

\hypersetup{
    colorlinks=true,
    linkcolor=blue,
    filecolor=magenta,
    urlcolor=cyan,
    citecolor=cyan
}

\begin{document}

\title{Testing non-local gravity through Ultra-Diffuse Galaxies kinematics}

\author{Filippo Bouchè}
\email{filippo.bouche-ssm@unina.it} 
\affiliation{Scuola Superiore Meridionale, Largo San Marcellino 10, 80138 Napoli, Italy}
\affiliation{Istituto Nazionale di Fisica Nucleare, Sez. di Napoli,  Via Cinthia 21, 80126 Napoli, Italy}

\author{Salvatore Capozziello}
\email{capozziello@na.infn.it}
\affiliation{Scuola Superiore Meridionale, Largo San Marcellino 10, 80138 Napoli, Italy}
\affiliation{Istituto Nazionale di Fisica Nucleare, Sez. di Napoli,  Via Cinthia 21, 80126 Napoli, Italy}
\affiliation{Dipartimento di Fisica ``E. Pancini'', Universit\`a  di Napoli  ``Federico II", Via Cinthia 21, 80126 Napoli, Italy}

\author{Ciro De Simone}
\email{ciro.desimone@unina.it}
\affiliation{Istituto Nazionale di Fisica Nucleare, Sez. di Napoli,  Via Cinthia 21, 80126 Napoli, Italy}
\affiliation{Dipartimento di Fisica ``E. Pancini'', Universit\`a  di Napoli  ``Federico II", Via Cinthia 21, 80126 Napoli, Italy}

\author{Vincenzo Salzano}
\email{vincenzo.salzano@usz.edu.pl}
\affiliation{Institute of Physics, University of Szczecin, Wielkopolska 15, 70-451 Szczecin, Poland}

\date{\today}

\begin{abstract}

The emergence of the Ultra-Diffuse Galaxies in recent years has posed a severe challenge to the galaxy formation models as well as the Extended Theories of Gravity. The existence of both dark matter lacking and dark matter dominated systems within the same family of astrophysical objects indeed requires the gravity models to be versatile enough to describe very different gravitational regimes. In this work, we study a non-local extension of the theory of General Relativity that has drawn increasing attention in recent years due to its capability to account for the late time cosmic acceleration without introducing any dark energy fluid. We leverage the kinematic data of three Ultra-Diffuse Galaxies: NGC 1052-DF2 and NGC 1052-DF4, which are dark matter lacking, and Dragonfly 44, which exhibits a highly dominant dark matter component. Our analysis shows that the non-local corrections to the Newtonian potential do not affect the kinematic predictions, hence no spoiling effects emerge when the Non-local Gravity model serves as a dark energy model. We additionally provide the minimum value that the characteristic non-local radii can reach at these mass scales.

\end{abstract}


\maketitle

\section{Introduction}

Since the first evidence of the ongoing acceleration of the Universe's expansion \cite{SupernovaCosmologyProject:1998vns, SupernovaSearchTeam:1998fmf, SupernovaSearchTeam:1998bnz}, the nature of the Dark Energy (DE) stands as one of the most puzzling issues in both cosmology and fundamental physics. A plethora of theoretical and phenomenological models has therefore been developed, ranging from a Cosmological Constant ($\Lambda$) associated to the quantum vacuum energy to modifications of gravity \cite{Copeland:2006wr, Bamba:2012cp, Nojiri:2017ncd, CANTATA:2021ktz}. 
However, considering the currently available data, both from the geometrical background and from the perspective of structures formation and evolution, the vast majority of these models does not provide any decisive answer to the issue. The hope is that the forthcoming IV-generation surveys (\textit{Euclid}\footnote{\url{https://www.euclid-ec.org/}}, Rubin Observatory\footnote{\url{https://www.lsst.org/}}, SKAO\footnote{\url{https://www.skao.int/}}, \textit{Roman Space Telescope}\footnote{\url{https://roman.gsfc.nasa.gov/}}, etc.) are expected to generate an extensive and groundbreaking collection of cosmological data, that may help to finally shed some light on the true nature of the DE. Discerning whether the DE is associated to some field or to a modification of the General Relativity (GR) as well as investigating the redshift-dependence of the DE equation of state \cite{DESI:2024mwx} are two of the primary scientific goals of the next decade.

A complementary approach for delving into the zoo of DE models relies on the analysis of the impact within the non-linear regime resulting from the introduction of extra Degrees of Freedom (DoF) in the theory. Therefore, the investigation of the dynamics of gravitationally bound systems emerges as the ideal framework to explore the feasibility of any modification of the standard $\Lambda$ Cold Dark Matter ($\Lambda$CDM) paradigm. Astrophysical data from a large variety of systems have been leveraged to test DE models in literature, ranging from the orbits of S2 star around Sgr A* \cite{Clifton:2020xhc, deMartino:2021daj, Borka:2021omc, Bahamonde:2021akc, DellaMonica:2021xcf, Jusufi:2022loj, Benisty:2023qcv, Zhang:2024fpm} up to the dynamical and gravitational features of galaxy groups \cite{Carlesi:2016yas, Gurzadyan:2017gii, Benisty:2019fzt, McLeod:2019cfg, Benisty:2023ofi, Benisty:2023vbz} and clusters \cite{Terukina:2013eqa, Wilcox:2015kna, Brax:2015lra, Pizzuti:2016ouw, Pizzuti:2017diz, Cataneo:2018mil, Laudato:2021mnm, Haridasu:2021hzq, Bouche:2022jts, Pizzuti:2021brr, Zamani:2024oep}. The dynamics of spiral \cite{Moffat:2014pia, Capozziello:2017rvz, Finch:2018gkh, Panpanich:2018cxo, deAlmeida:2018kwq, Petersen:2020vks, Chae:2022oft}, elliptical \cite{Rodrigues:2014xka, Capozziello:2020dvd1, Borka:2022vgi} and Ultra-Diffuse Galaxies (UDGs) \cite{Moffat:2018pab, Haghi:2019zpc, Islam:2019szu, Laudato:2022vmq, Laudato:2022drz, Bhatia:2023pts} has been analysed as well.

In this article, we focus on UDGs whose internal kinematic can be traced by globular clusters. UDGs are a specific class of low surface brightness galaxies, characterized by either the lack of star-forming gas \cite{vanDokkum:2014cea} or a very low star formation efficiency \cite{Leisman_2017_UDG}. UDGs also show remarkable similarities with dwarf ellipticals and dwarf spheroidal galaxies, suggesting that they might be part of the same population of galaxies with different size \cite{Conselice_2018_UDG}. Nevertheless, some of these systems present an unexpectedly large number of globular clusters \cite{Beasley_2016_GCs, Peng_2016_GCs, Amorisco_2018_GCs} that enables an accurate reconstruction of their internal galactic velocity dispersion. UDGs have been recently found in very different environments, ranging from extremely under-dense voids \cite{Rom_n_2019_UDG_voids} up to galaxy groups \cite{Merritt_2016_UDG_group, Smith_Castelli_2016_UDG_group}, clusters \cite{van_der_Burg_2016_UDG_cluster} and filaments \cite{Mart_nez_Delgado_2016_UDG_filament}. Moreover, although these galaxies share common stellar features, they exhibit extremely different DM compositions, with both completely DM-dominated and baryon-dominated UDGs that have been observed \cite{van_Dokkum_2018_ml, vanDokkum:2019fdc}.

Even though the existence of DM-lacking UDGs strongly challenges those Extended Theories of Gravity (ETGs) that aim at replacing (or at least reducing) DM and its effects with geometrical or non-particle-based sources, the identification of these galaxies in a wide variety of environments as well as their various compositions render UDGs highly attractive exactly for testing such theories of gravity. On the one hand, the environmental effects can be disentangled from the imprints of the theories; on the other hand, it is possible to analyse very different regimes within the same family of astrophysical objects. In this paper, we therefore use kinematic data from three different UDGs to carry out a stress test of a DE model based upon a non-local extension of GR. Our dataset comprises precise estimates of the velocity dispersion for many globular clusters associated to UDGs \cite{van_Dokkum_2018_ml, van_Dokkum_2019_df4, vanDokkum:2019fdc}. Specifically, we analyse two DM-lacking galaxies (NGC 1052-DF2 and NGC 1052-DF4) and one galaxy that is DM dominated (Dragonfly 44).

In the vast arena of ETGs, in this work we will focus our attention on the non-local gravity theories. Inspired by the Effective Field Theories in the quantum realm, these models have recently drawn increasing attention due to their effectiveness to improve the behaviour of the gravitational interaction both in the ultraviolet (UV) and infrared (IR) regime \cite{Capozziello:2021krv}. Carrying on the work that we started in \cite{Bouche:2022jts}, we therefore investigate the astrophysical features of the non-local model proposed in \cite{Deser:2007jk}, which accounts for the cosmic acceleration through a delayed response to the transition from the radiation to the matter cosmic epoch.

This article is organized as follows: in Sec.~\ref{sec_theory} we present the non-local theoretical scenario which we are going to test, and we highlight its main characteristic features. In Sec.~\ref{sec_UDG_theory}, we describe the physical quantities related to the UDGs kinematic which we have analyzed, and how we model UDGs internal structure. Additionally, in Sec.~\ref{sec_UDG_data} we overview the properties of the three UDGs that have been investigated throughout this work, the associated datasets that we used for our analysis are presented, and the main ingredients of our statistical analysis. Results are discussed in Sec.~\ref{sec:results}. Finally, we draw our conclusions in Sec.~\ref{sec:conclusions}.

\section{Non-local gravity}\label{sec_theory}

The Quantum Field Theory (QFT) typically exhibits non-local features associated to the one-loop effective actions. The non-locality indicates that we are approaching an energy scale where the effective theory becomes inadequate, thus necessitating the introduction of additional fields \cite{Capozziello:2021krv}. Accordingly, incorporating non-local corrections into the Hilbert-Einstein action offers a quantum-inspired effective strategy to improve the behaviour of the gravitational interaction, addressing potential breakdowns of the General Relativity in both the UV and IR regimes. Following this approach, many non-local extensions of GR have been proposed in literature \cite{Wetterich:1997bz, Deser:2007jk, Biswas:2011ar, Briscese:2012ys, Maggiore:2013mea, Maggiore:2014sia, Deser:2019lmm}. These can be grouped in two families: the Infinite Derivatives theories of Gravity (IDGs), whose non-local operators are made of entire analytic functions of a differential operator; and the Integral Kernel theories of Gravity (IKGs), based on non-local operators that are characterized by transcendental functions of the geometric fields. The former ones incorporate short-range non-localities arising from the derivative nature of non-local operators, hence tackling the UV shortcomings of GR, such as the loss of unitarity caused by ghosts \cite{Biswas:2011ar, Briscese:2012ys, Buoninfante:2020ctr} and the existence of singularities \cite{Biswas:2005qr, Buoninfante:2018xiw, Buoninfante:2018xif}. The latter ones give rise to non-localities on large scales due to the integral nature of IKG operators, and are therefore applied for dealing with the IR issues of GR, e.g. the late time cosmic acceleration \cite{Deffayet:2009ca,Belgacem:2020pdz, Capozziello:2023ccw}.

In this paper, we focus on the metric IKG proposed in \cite{Deser:2007jk}, which is equivalent to the generalized model of \cite{Bahamonde:2017sdo} if the coupling constants are set to $1$ and $-1$. The examined non-local model is characterized by a distortion function, $f(\Box^{-1}R)$, that extends the standard GR action
\begin{equation}\label{DWaction}
    S = \frac{1}{16 \pi G} \int d^{4}x \sqrt{-g} \,\Big\{ R \big[ 1 + f(\Box^{-1}R) \big] \Big\} \, .
\end{equation}
The non-local operator can be rewritten as the Green function associated to the differential operator, as any typical IKG operator,
\begin{equation}\label{InverseBox}
        \Box^{-1} R(x) \equiv \int d^{4}x' G(x,x') R(x') \, .
\end{equation}
This kind of effective non-local corrections naturally emerge in the quantum realm, either when non-perturbative methods are applied to the dimensional regularization of QFTs on curved spacetime \cite{Barvinsky:2014lja}, or the regularization of quantum black holes \cite{Knorr:2022kqp} as well as the quantum conformal anomalies \cite{Coriano:2017mux} are considered.

To carry out our analysis, we have recast the action in Eq.~(\ref{DWaction}) as a scalar-tensor theory,
\begin{equation}\label{STaction}
    S = \frac{1}{16 \pi G} \int d^{4}x \sqrt{-g} \, \Big[ R \big( 1 + f(\eta) - \xi \big) - \nabla^{\mu}\xi \, \nabla_{\mu}\eta \Big] \, ,
\end{equation}
as done in \cite{Nojiri:2007uq}. The two auxiliary scalar fields are described by the following equations of motion
\begin{equation}
    \Box \eta = R \; , \quad\quad \Box \xi = -R \, \frac{df(\eta)}{d\eta} \, ,
\end{equation}
where the equation for $\eta$ simply reduces to the constraint $\eta = \Box^{-1}R$. Moreover, the gravitational field equations read
\begin{align}
    G_{\mu\nu} \;=\; &\frac{1}{1+f(\eta)-\xi} \, \bigg\{ ( 8\pi G ) \, T_{\mu\nu}^{(m)} - \frac{1}{2} g_{\mu\nu} \, \partial^{\alpha}\xi \, \partial_{\alpha}\eta \nonumber \\
    &+ \frac{1}{2} \Big( \partial_{\mu}\xi \, \partial_{\nu}\eta + \partial_{\mu}\eta \, \partial_{\nu}\xi \Big) - \Big( g_{\mu\nu} \square - \nabla_{\mu} \nabla_{\nu} \Big) \nonumber \\
    &\times \Big[ f(\eta) - \xi \Big] \bigg\} \, ,
\end{align}
where the non-local geometric terms can be moved to the right hand side, since the field equations fulfill the Bianchi's identities \cite{Deser:2007jk}.
The main phenomenological feature of this non-local model is its capability to account for the late time cosmic acceleration without introducing any additional DE fluid. If the action in Eq.~(\ref{DWaction}) is written in terms of the Friedman-Lema\^itre-Robertson-Walker (FLRW) metric and the initial conditions are such that the non-local correction vanishes during the radiation epoch of the Universe, then the distortion function yields a delayed response to the radiation-to-matter dominance transition that drives the onset of the cosmic acceleration at the late epoch \cite{Deser:2007jk}. The non-local model is therefore able to exactly reproduce the $\Lambda$CDM cosmic expansion history \cite{Deffayet:2009ca}. Being the background evolution equivalent to that of the concordance model, the only way to test the non-local model under consideration is to investigate how the non-local gravitational corrections impact the cosmic structures. The growth of perturbations has been analysed in \cite{Amendola:2019fhc} through a joint analysis of the Cosmic Microwave Background radiation, Redshift Space Distortion data, and the SDSS Supernovae Ia catalogue. This analysis shows a lower value of the amplitude of matter perturbations, $\sigma_8$, with respect to the $\Lambda$CDM model, thus alleviating the so-called (growth) $\sigma_8$-tension \cite{Bouche:2023xjw}. 

Moreover, astrophysical tests have been carried out to test the weak field regime: the orbits of the star S2 around Sgr A* \cite{Borka:2021omc} as well as the galaxy cluster lensing \cite{Bouche:2022jts} have been analysed in the non-local gravity framework. The model has shown no spoiling effects across the different scales. 

In this paper, we present a new astrophysical test of the non-local model, which has been carried out at two unexplored mass scales: $\sim 10^8 M_\odot$, which corresponds to the DM lacking UDGs, and $\sim 10^{11} M_\odot$, associated to the DM dominated UDG. To derive the non-local theoretical predictions for the velocity dispersion, we have relied on the Newtonian (point-mass) potential
\begin{align}\label{phi}
    \phi(r) &= \phi_{\text{GR}}(r) + \phi_{\text{NL}} (r)\, , \\
    \phi_{\text{GR}}(r) &= - \frac{GM}{r} \, , \nonumber \\
    \phi_{\text{NL}}(r) &=  \frac{G^{2}M^{2}}{2c^{2}r^{2}} \Bigg[ \frac{14}{9} + \frac{18 r_{\xi} - 11 r_{\eta}}{6 r_{\eta} r_{\xi}} \, r \Bigg] \, , \nonumber
\end{align}
which has been derived in \cite{Dialektopoulos:2018iph}. This potential has been derived in the Post-Newtonian approximation and thus describe systems that are characterized by weak gravity and slow motion. This approximation has been also tested against numerical relativity showing extremely accurate results even beyond its nominal regime of validity \cite{Will:2011nz}. Therefore, the weak field potential in Eq.~(\ref{phi}) has been effectively applied to describe the motion of the star S2 around the galactic center \cite{Dialektopoulos:2018iph}. Throughout this paper, we safely use the potential to describe the motion of Globular Clusters within the UDGs gravitational potential well. It is worth noticing that the non-local Newtonian potential shows two new length scales, $r_\eta$ and $r_\xi$, that arise in the $\mathcal{O}(4)$ order of the Post-Newtonian expansion, and are associated to the scalar degrees of freedom of Eq.~(\ref{STaction}). Since the scalar fields $\eta$ and $\xi$ encode the non-local correction to the Hilbert-Einstein action, the two radii $r_\eta$ and $r_\xi$ can be interpreted as the characteristic interaction lengths associated to the gravitational non-localities.  The form of the potential in Eq.~(\ref{phi}) is associated to an exponential form of the distortion function,
\begin{equation}
    f(\eta) = 1 + e^{\eta} \, ,
\end{equation}
which is obtained by requiring that Eq.~(\ref{STaction}) is invariant under point transformations in a spherically symmetric background spacetime. This specific form of the distortion function is thus associated to the existence of Noether symmetries \cite{Dialektopoulos:2018qoe}.

As a final remark with regard to non-local gravity, it is important to notice that, as any extension of GR, also the non-local model that we have described need to be tested against the observations at the well-tested Solar System scale. In \cite{Deser:2013uya} it has been proposed a simple screening mechanism, which is based on a change of sign of the non-local operator inside gravitationally bound systems. As a consequence, an Heaviside step function can be incorporated in $f(\Box^{-1}R)$ by taking advantage of the freedom in choosing the distortion function. Therefore, the non-local effects would vanish in bound objects. However, in \cite{Belgacem:2018wtb} it has been shown that the value of $\,\Box^{-1}R$ is actually negative also at Solar System scale, and this procedure is thereby unfeasible. It may follow a time dependence of the effective Newton constant in the small scale limit, hence the non-local model would be ruled out by Lunar Laser Ranging observations. The failure of this perfect screening mechanism makes room for testing the non-local model by leveraging astrophysical data from gravitationally bound cosmic structures. However, since the lack of a proper screening at Solar System scale is a critical issue, two further considerations are deserved before drawing such a strong conclusion as the one of \cite{Belgacem:2018wtb}. On the one hand, it is still uncertain how FLRW background quantities behave when evaluated at Solar System scales, where the system strongly decouples from the Hubble flow. A full non-linear solution that accounts for both time and scale dependencies would be actually needed. On the other hand, the most popular screening mechanisms \cite{Vainshtein:1972sx, Capozziello:2007eu, Babichev:2009ee, Hinterbichler:2010es}, which are used in ETGs, rely on a scalar field coupled to matter, and mediating the additional gravitational effects. For high densities, the coupling would be suppressed, hence no deviation from GR. For lower densities, the scalar field coupling would be enhanced and the modifications to GR should become effective. Since the non-local terms result in effective scalar fields depending on the scale \cite{Acunzo:2021gqc}, these screening mechanisms could be suitable for non-local gravity as well. Since this is a crucial topic for the non-locally ETGs, further studies in this direction are necessary in the future.

\section{Galactic kinematics}\label{sec_UDG_theory}

The internal kinematic of UDGs can be studied through the analysis of the velocity dispersion as a function of the galactic radius. This task requires the solution of the Jeans equation, which reads
\begin{equation}\label{JeansEq}
    \frac{d}{dr}\big[ \ell(r) \sigma^2_r(r) \big] + 2 \frac{\beta(r)}{r} \ell(r) \sigma^2_r(r) = -\ell(r) \frac{G M(r)}{r^2} \, ,
\end{equation}
where both the spherical symmetry and no-streaming motions have been assumed \cite{Mamon:2004xk}. Here, $\ell(r)$ is the luminosity density of the galaxy, $\sigma_r(r)$ is the radial velocity dispersion, and $\beta(r)$ is the anisotropy parameter defined as
\begin{equation}
    \beta(r) = 1 - \frac{\sigma^2_t(r)}{\sigma^2_r(r)} \, ,
\end{equation}
with $\sigma_t(r)$ being the tangential velocity dispersion. Modeling the internal kinematics of spherically symmetric, slowly rotating galaxies poses a severe challenge due to the fact that Eq.~(\ref{JeansEq}) involves two unknown radial functions: the mass distribution and the anisotropy parameter. As a consequence of this degeneracy, strong assumptions on the functional form of $\beta(r)$ are needed. Throughout this paper we consider two possible anisotropy profiles,
\begin{equation}\label{anisotropyProfile}
\begin{aligned}
    \beta(r) &= \beta_c \, , \\
    \beta(r) &= \beta_0 + (\beta_\infty - \beta_0 ) \frac{r}{r + r_a} \, ,
\end{aligned}
\end{equation}
respectively, a constant \cite{Napolitano:2008ft} and a radial profile, inspired by the kinematic reconstruction in UDGs shown in \cite{Zhang:2015pca}. 

The Jeans equation, Eq.~(\ref{JeansEq}), has a general solution
\begin{equation}\label{EQveldisp}
    \ell(r) \sigma^2_r(r) = \frac{1}{f(r)} \int^\infty_r ds \, f(s) \ell(s) \frac{GM(s)}{s^2} \, ,
\end{equation}
where the function $f(r)$ depends on the anisotropy profile
\begin{equation}
    \frac{d \ln f(r)}{d \ln r} = 2 \beta(r) \, .
\end{equation}
The projection of the velocity dispersion in Eq.~(\ref{EQveldisp}) along the line-of-sight, $\sigma^2_{los}$, which is actually the quantity that we measure, yields \cite{10.1093/mnras/200.2.361}
\begin{align}\label{sigmaLOS}
    \sigma^2_{los}(R) =& \frac{2}{I(R)}  \bigg[ \int_R^\infty dr\, \frac{r}{\sqrt{r^2 - R^2}} \ell(r) \sigma^2_r(r) \\
    &- R^2 \int_R^\infty dr\, \frac{\beta(r)}{r \sqrt{r^2 - R^2}} \ell(r) \sigma^2_r(r) \bigg] \, , \nonumber
\end{align}
where $R$ is the 2D projected radius, and $I(R)$ is the surface density of the galaxy. For our choice of the anisotropy profiles, Eq.~(\ref{sigmaLOS}) can be rewritten as
\begin{equation}\label{theoPred}
    \sigma^2_{los}(R) = \frac{2G}{I(R)} \int_R^\infty dr \, K \Big( \frac{r}{R}, \frac{r_a}{R} \Big) \ell(r) \frac{M(r)}{r} \, ,
\end{equation}
where the functional form of the kernel $K(x,y)$ is defined for few simple choices of $\beta(r)$ \cite{Mamon:2004xk}, including those ones introduced above.

Since it is always possible to define an effective mass for any ETG, Eq.~(\ref{phi}) can be recast as
\begin{equation}\label{eff_mass}
    M_{\text{eff}}(r) = \frac{r^2}{G} \frac{d \Phi(r)}{dr} = M_{\text{GR}}(r) + M_{\text{NL}}(r) \, ,
\end{equation}
where $M_{\text{GR}}$ is the contribution to the effective mass from $\phi_{\text{GR}}$ in Eq.~(\ref{phi}), while $M_{\text{NL}}$ is the non-local correction coming from $\phi_{\text{NL}}$ in Eq.~(\ref{phi}). As a consequence, the non-local theoretical predictions for the velocity dispersion can be derived from Eq.~(\ref{theoPred}). Considering that the potential in Eq.~(\ref{phi}) is a point-mass potential, it has to be generalized for extended mass distributions, finally having:
\begin{align}\label{ext_PhiGR}
    \Phi_{\text{GR}}(r) &= \Phi^{\text{int}}_{\text{GR}}(r) + \Phi^{\text{ext}}_{\text{GR}}(r)\, , \\
    \Phi^{\text{int}}_{\text{GR}}(r) &= \int_{0}^{r} dx \, \rho(x) x^2 \int_{0}^{\pi} d\theta \, \sin \theta  \int_{0}^{2\pi} d\varphi \, \phi_{\text{GR}}(x)\, , \nonumber \\
    \Phi^{\text{ext}}_{\text{GR}}(r) &= \int_{r}^{\infty} dx \, \rho(x) x^2 \int_{0}^{\pi} d\theta \, \sin \theta  \int_{0}^{2\pi} d\varphi \, \phi_{\text{GR}}(x) \, , \nonumber 
\end{align}
and
\begin{align}\label{ext_PhiNL}
    \Phi_{\text{NL}}(r) &= \Phi^{\text{int}}_{\text{NL}}(r) + \Phi^{\text{ext}}_{\text{NL}}(r)\, , \\
    \Phi^{\text{int}}_{\text{NL}}(r) &= 2 \, \bigg[ 
    \int_{0}^{r} dx \, \rho(x) x^2 \int_{0}^{x} dy \, \rho(y) y^2 \int_{0}^{\pi} d\theta \sin \theta \nonumber \\ 
    &\times \int_{0}^{\pi} d\theta' \sin \theta' \int_{0}^{2\pi} d\varphi \int_{0}^{2\pi} d\varphi' \Phi_{\text{NL}}(y)\, , \nonumber \\
    \Phi^{\text{ext}}_{\text{NL}}(r) &= 2 \, \bigg[
    \int_{r}^{\infty} dx \, \rho(x) x^2 \int_{x}^{\infty} dy \, \rho(y) y^2 \int_{0}^{\pi} d\theta \sin \theta \nonumber \\ 
    &\times \int_{0}^{\pi} d\theta' \sin \theta' \int_{0}^{2\pi} d\varphi \int_{0}^{2\pi} d\varphi' \Phi_{\text{NL}}(y) \bigg] \,. \nonumber
\end{align}
It's worth to stress that the extension of the non-local correction to the Newtonian potential involves a $M^2$ term. Since the mass element can be written as
\begin{equation}
dM = \rho(x') \, x'^{2} \, dx' \, \sin \theta \, d\theta \, d\phi\, ,
\end{equation}
and considering that $dM^2 = 2 M dM$, with
\begin{equation}\label{2.27}
M(r) = \, 4\pi \int_{0}^{r} dx' \, x'^{2} \, \rho(x')\, 
\end{equation}
we finally get Eq.~(\ref{ext_PhiNL}). Here, $r$ is defined as
\begin{equation}
r = |\boldsymbol{x} - \boldsymbol{x'}| = \big(x^{2}+ x'^{2} - 2 \, x \, x' \, \mathrm{cos} \theta \big)^{1/2}\, ,
\end{equation}
where $\boldsymbol{x}$ is the vector position of the point in the space where we want to calculate the potential, and $\boldsymbol{x'}$ is the vector position connected to the mass distribution (source of the gravitational potential). The integration over the radial coordinate has to be performed on both the intervals $[0,r]$ and $[r,\infty)$, because Newton's theorems are not guaranteed in non-local gravity, hence the Gauss theorem does not hold\footnote{As discussed in \cite{Capozziello_GaussTheo}, the violation of the Gauss theorem in the ETGs is not an issue, since the conservation laws are guaranteed by the Bianchi identities.}. 

It is worth noticing that 
we could not carry out the integration over the whole radial range. Instead, we had to set a cutoff to account for the slow convergence of the matter density profiles that we selected to model the UDGs. A cutoff value of $r_c=50$ kpc has been established, corresponding to $\approx5$ times the fixed radius that has been used in \cite{vanDokkum:2019fdc} to extract a reliable estimate of the UDG dynamical mass. 
We have thus further checked the impact of our cutoff choice on the parameters inference by carrying out an additional analysis for the fiducial non-local cases, in which we lower the cutoff to the value used in \cite{vanDokkum:2019fdc}, $r_c=10$ kpc. Eventually, we recover statistically equivalent values for the whole set of parameters but the non-local length scales. Nevertheless, this change is easily predictable when looking at the expression of the potential and its nonlocal correction, Eq.~(\ref{phi}). Although it is not possible to obtain an analytical relation among $r_{c}$, $M(r_{c})$ and the nonlocal radii, one can easily verify that lowering both the cutoff distance and the mass content requires a corresponding lowering in the nonlocal radii.

\subsection{Galactic internal structure}\label{sec_2C}

The appropriate matter density profiles have to be selected in order to properly model the internal structure of UDGs and extend the Newtonian potential accordingly. Throughout this paper we use the generalized Navarro-Frenk-White (gNFW) profile \cite{Zhao:1995cp} to describe the dark matter component of the UDGs,
\begin{equation}\label{gNFW}
    \rho_{\text{gNFW}}(r) = \rho_s \bigg(\frac{r}{r_s}\bigg)^{-\gamma} \bigg( 1 + \frac{r}{r_s} \bigg)^{\gamma - 3} \, ,
\end{equation}
where $\rho_s$ and $r_s$ are the characteristic density and radius of the profile, while $\gamma$ modulates the inner slope thus also allowing a cored profile \cite{Moore:1999gc, Brook:2021veq}. This dark matter profile can be expressed in terms of the set of free parameters $\{ c_{200}, M_{200}, \gamma \}$, defined as:
\begin{align}
    M_{\Delta} &= \Delta \rho_{\text{c}} \, \frac{4 \pi r_{\Delta}^3}{3} \, , \\
    r_s &= \frac{r_{\Delta}}{c_{\Delta}} \, ,\\
    \rho_s &= \Delta \rho_{\text{c}} \, \frac{(3-\gamma) c_{\Delta}^{\gamma}}{3 \, _{2}F_{1}[3-\gamma, 3-\gamma, 4-\gamma, - c_{\Delta}]} \, .
\end{align}
Here, $\rho_{\text{c}}(z)$ is the critical density of the Universe at the UDG redshift, $_{2}F_{1}$ is the hypergeometric function, and we selected an overdensity value $\Delta = 200$, since $M_{200}$ is a good estimate of the virial mass of the galaxies according to the spherical collapse model.

For what concerns the stellar component, we model it through the Sérsic optical surface brightness profile \cite{1968adga.book.S, Graham:2005fy},
\begin{equation}\label{Sersic}
    I(R) = I_0 \exp \bigg[ - \bigg( \frac{R}{a_s} \bigg)^{\frac{1}{n}} \bigg] \, ,
\end{equation}
where $I_0$ is the central surface brightness of the galaxy
\begin{equation}
    I_0 = \frac{L_{\text{tot}}}{2 \pi \, n \, a_s^2 \, \Gamma(2 n)} \, ,
\end{equation}
and $L_{\text{tot}}$ is the total luminosity of the galaxy,
\begin{equation}
    L _{\text{tot}} = 10^{-0.4 [m_{V_{606}} - \mu(D) - M_{\odot , V_{606}}]} \,\, .
\end{equation}
Here, $m_{V_{606}}$ is the apparent magnitude of the galaxy, $M_{\odot , V_{606}} = 4.72$ is the Sun’s absolute magnitude in the $V_{606}$ photometric band \cite{Willmer_2018}, and $\mu(D)= 5 \log_{10}(D) +25$ is the distance modulus, where $D$ is the distance of the galaxy in Mpc. The characteristic parameters of the Sérsic profile are the Sérsic index, $n$, which regulates the shape of the profile, and the Sérsic scale parameter, $a_s$, that can be recast as \cite{Graham:2003dg}
\begin{equation}
    a_s = \frac{R_{\text{eff}}}{(b_n)^n} \, ,
\end{equation}
with $R_{\text{eff}}$ being the half-light projected radius, and $b_n = 2n-0.33$. The deprojection of the Sérsic profile in Eq.~(\ref{Sersic}) yields the luminosity density profile \cite{1997A&A.321.111P}
\begin{equation}
    \rho_*(r) = \Upsilon^* \ell (r) = \Upsilon^* \ell_1 \, \Tilde{\ell}\bigg( \frac{r}{a_s} \bigg) \, ,
\end{equation}
where
\begin{equation}
\begin{aligned}
    \ell_1 &= \frac{L_{\text{tot}}}{4 \pi \, n \, a_s^3} \, \frac{1}{\Gamma[(3-p)n]} \, , \\
    \Tilde{\ell}(x) &= x^{-p(n)} \exp \big( -x^{1/n} \big) \, .
\end{aligned}
\end{equation}
An approximate expression for the function $p(n)$ is provided by \cite{Neto:1999gx}:
\begin{equation}
    p(n) \simeq  1.0 - 0.6097/n + 0.05463/n^2 \, .
\end{equation}

\section{Data}\label{sec_UDG_data}

In this work, we have tested the theoretical predictions of the non-local model against the observations of the velocity dispersion of three UDGs: NGC 1052-DF2, NGC 1052-DF4 and Dragonfly 44. 

The former was first observed by using the Dragonfly Telescope Array \cite{Abraham:2014lfa}. Then, the structural properties have been further investigated through photometric observations with the \textit{Hubble Space Telescope} \cite{van_Dokkum_2018_ml2}, while the spectroscopy of the NGC 1052-DF2 globular clusters has been carried out with the twin telescopes of the W. M. Keck Observatory \cite{van_Dokkum_2018_ml}. This UDG is characterized by a Sérsic index $n=0.6$, an effective radius $R_{\text{eff}}=22.6$ arcsec, and a central surface brightness in the $V_{606}$ band $I_0 = 24.4$ mag arcsec$^{-2}$. Given the fiducial distance $D=20$ Mpc \cite{van_Dokkum_2018_ml}, the observed $I_0$ corresponds to an absolute magnitude $M_{V_{606}} = -15.4$. NGC 1052-DF2 also presents a partial ellipticity with an axis ratio $b/a=0.85$. Regarding the radial velocity and the velocity dispersion of the ten globular clusters that we have used for our statistical analysis\footnote{The data provided by \cite{van_Dokkum_2018_ml} can be found at \url{http://www.astro.yale.edu/dokkum/outgoing/ascii_table.txt}.}, it is worth noticing that one of the velocities has been revised in \cite{vandokkum2018revised}.

For what concerns NGC 1052-DF4, it belongs to the same group of the previous galaxy and also shows similar structural properties. The redshift of the galaxy group NGC 1052 is $z = 0.004963$, as reported in the NED database \footnote{The NASA/IPAC Extragalactic Database (NED) is funded by the National Aeronautics and Space Administration and operated by the California Institute of Technology.}.
The observations of NGC 1052-DF4 were realized by using the Low Resolution Imaging Spectrograph on the Keck I telescope, and reported the presence of seven luminous globular clusters extending out to $7$ kpc from the centre of the galaxy \cite{van_Dokkum_2019_df4}. The data that we have used for the statistical analysis are reported in Table 1 of \cite{van_Dokkum_2019_df4}. NGC 1052-DF4 can be described through a Sérsic profile with $n=0.79$ and $R_{\text{eff}}=1.6$ kpc at the fiducial distance $D=20$ Mpc, hence $R_{\text{eff}}=16.58$ arcsec. The observed central surface brightness is $I_0 = 23.7$ mag arcsec$^{-2}$, with a corresponding absolute magnitude $M_{V_{606}} = -15.0$. Moreover, the galaxy is characterized by a partial ellipticity with an axis ratio $b/a = 0.89$.

Finally, the spectroscopic analysis of Dragonfly 44 has been carried out through the Keck Cosmic Web Imager on the Keck II Telescope \cite{vanDokkum:2019fdc}. This galaxy is located in the Coma cluster, whose redshift is $z = 0.023156$ as reported by the SIMBAD catalogue \cite{Wenger:2000sw}. Consequently, the estimated distance is $D=102 \pm 14$ Mpc \cite{Thomsen:1997vz}. The stellar profile is characterized by a Sérsic index $n=0.94$ and an effective radius $R_{\text{eff}}=4.7$ kpc at $D=100$ Mpc, corresponding to $R_{\text{eff}}=16.58$ arcsec. The central surface brightness of the galaxy is $I_0 = 24.1$ mag arcsec$^{-2}$, such that the absolute magnitude in the $V_{606}$ band is $M_{V_{606}} = -16.2$. Dragonfly 44 also shows an axis ratio $b/a = 0.68$. The kinematic data that we have used in our analysis are reported in Table 2 of \cite{vanDokkum:2019fdc}, where the radial circular velocities and the dispersion are given in nine radial bins.

\subsection{Statistical Analysis}\label{sec:statistical}

The aim of this work is to constrain the parameters of the non-local model defined in Eq.~(\ref{DWaction}) as well as the astrophysical parameters describing both the internal structure and the kinematics of the UDGs under investigation. Assuming a Gaussian likelihood, the $\chi^2$ function reads
\begin{equation}
    \chi^2(v_{sys}, \boldsymbol{\theta}) = \sum_{i=1}^{\mathcal{N}_{data}} \bigg\{ \frac{(v_i - v_{sys})^2}{\sigma^2_i(\boldsymbol{\theta})} + \ln \big[ 2\pi \sigma^2_i(\boldsymbol{\theta}) \big] \bigg\} \, ,
\end{equation}
for NGC 1052-DF2 and DF4, which are characterized by an high systemic velocity, $v_{sys}$ \cite{van_Dokkum_2018_ml, van_Dokkum_2019_df4}. Here, $\sigma^2_i(\boldsymbol{\theta}) = \sigma^2_{los, \, i}(\boldsymbol{\theta}) + \sigma^2_{v_i}$, with $\sigma_{v_{i}}$ the measurement uncertainties, hence the normalization factor of the Gaussian distribution depends on the free parameters. For what concerns Dragonfly 44, whose systemic velocity is consistent with zero \cite{vanDokkum:2019fdc}, the $\chi^2$ function instead reads
\begin{equation}
    \chi^2(\boldsymbol{\theta}) = \sum_{i=1}^{\mathcal{N}_{data}} \bigg[ \frac{\sigma_{\text{eff}, \, i} - \sigma_{los, \, i}(\boldsymbol{\theta})}{\delta \sigma_{\text{eff}, \, i}} \bigg]^2 \, ,
\end{equation}
where $\sigma^2_{\text{eff}, \, i} = \sigma^2_i + v^2_i$ is the effective velocity dispersion, and $\delta \sigma_{\text{eff}, \, i}$ is the corresponding uncertainty. Note that we have always imposed a requirements while computing the line-of-sight velocity dispersion, asking $\sigma^2_{los, \, i}(\boldsymbol{\theta})$ to be real-valued. 

Throughout our analysis we have assumed the prior distributions detailed in Tab.~\ref{tab:priors}. It is worth noticing that we used a piece of additional information about the internal structure of the UDGs in order to ensure the exploration of a physically reasonable region of the gNFW parameter space, due to the fact that the expected extension of the dark matter halo goes far beyond the range of distances provided by the observation. In particular, the relation $c_{200} = f_c(M_{200})$ between the gNFW concentration and mass parameters has been used in all the examined scenarios that involved a dark matter component. The specific functional form for $f_{c}$ is derived in \cite{Correa:2015dva} through a semi-analytic model that combines an analytic description of the halo mass accretion history with an empirical relation between the halo concentration and its formation time, obtained by fitting the results of N-body simulations. 

Moreover, we carried out our analysis both with and without an additional relation relating the stellar mass $M_{\ast} = \Upsilon_{\ast} L_{\text{tot}}$ with the expected (cor-)relation between stellar and dark matter mass, $M_{\ast} = f_M(M_{200})$, provided by \cite{Rodr_guez_Puebla_2017}. This stellar-to-halo mass relation (SHMR), built through a semi-empirical model, describes the co-evolution of galaxies and host halos across a wide redshift range.

{\renewcommand{\tabcolsep}{2.5mm}
{\renewcommand{\arraystretch}{2.}
\begin{table}[h]
\begin{minipage}{8.5cm}
\huge
\centering
\caption{Prior distributions used throughout our statistical analysis. $\mathcal{N}$ refers to normal distributions, $\log\mathcal{N}$ to Gaussian priors in the log-variable, and $\mathcal{U}$ to flat priors. For the parameters $v_{sys}$, $D$ and $\Upsilon_{\ast}$ we also enfore the positiveness condition. Moreover, $f_c$ denotes the function $f_c(M_{200})$ from the $c-M$ relation of \cite{Correa:2015dva}, and $f_M$ denotes the function $f_M(D,\Upsilon_*)$ from the SHMR of \cite{Rodr_guez_Puebla_2017}. Details on these relations are given in Sec.~\ref{sec:statistical}.}\label{tab:priors} 
\vspace{5pt}
\resizebox*{0.85\textwidth}{!}{
\begin{tabular}{c|cc}
\hline
Parameters & Priors & References \\
\hline
\multirow{2}{*}{$v_{sys}$ (km s$^{-1}$)} & DF2: $\mathcal{N}(1801.6, \, 5.0)$ & \cite{Wasserman_2018_df2} \\ & DF4: $\mathcal{N}(1444.60, \, 7.75)$ & \cite{van_Dokkum_2019_df4} \\ 
\hline
\multirow{3}{*}{$D$ (Mpc)} & DF2: $\mathcal{N}(22.1, \, 1.2)$ & \cite{Shen_2021_dist} \\ & DF4: $\mathcal{N}(22.1, \, 1.2)$ & \cite{Shen_2021_dist} \\ & DF44: $\mathcal{N}(102, \, 14)$ & \cite{Thomsen:1997vz} \\ 
\hline
\multirow{3}{*}{$\Upsilon_*$} & DF2: $\mathcal{N}(1.7, \, 0.5)$ & \cite{Wasserman_2018_df2} \\ & DF4: $\mathcal{N}(2.0, \, 0.5)$ & \cite{van_Dokkum_2019_df4} \\ & DF44: $\log\mathcal{N}(1.5, \, 0.1)$ & \cite{Wasserman_2019} \\ 
\hline
$1-\beta_i$ & $\log\mathcal{N}(0, \, 0.5) \in (-10,1)$ & \cite{Wasserman_2018_df2} \\ 
\hline
$r_a$ (kpc) & $\mathcal{U}(0, \, \infty)$ & $-$ \\
\hline
$c_{200}$ & $\log\mathcal{N}(f_c\, , \, 0.16) \in (0,40)$ & \cite{Correa:2015dva} \\ 
\hline
$M_{200}$ & $\mathcal{U}(0, \, 20)$ & $-$ \\
\hline
SHMR: $\log M_{\ast}$ & $\log\mathcal{N}(f_M\, , \, 0.3)$ & \cite{Rodr_guez_Puebla_2017} \\
\hline
$\gamma$ & $\mathcal{U}(0, \, 2.2)$ & $-$ \\
\hline
$\log r_{\eta}$ (kpc) & $\mathcal{U}(-10, \, 10)$ & $-$ \\
\hline
$\log r_{\xi}$ (kpc) & $\mathcal{U}(-10, \, 10)$ & $-$ \\
\hline
\end{tabular}}
\end{minipage}
\end{table}}}

Given the prior and the likelihood distributions, we carried out our inference through a Monte Carlo Markov Chain (MCMC) analysis. We made use of the $\mathtt{emcee}$ package \cite{emcee}, which implements the affine invariant ensemble sampler by \cite{ensamble_sampler} and provides a convergence analysis tool based on the integrated autocorrelation time \cite{Sokal1997}. Moreover, to assess the validity of the non-local model against the standard GR scenario, we have computed the Bayes factor, $\mathcal{B}^{\,i}_j$, where $\mathcal{M}_{j}$ is the reference model against which the model $\mathcal{M}_{i}$ is tested. The Bayes factor is defined as the ratio of the Bayesian evidence of the two models, which have been calculated through the nested sampling algorithm proposed in \cite{Mukherjee:2005wg}. Since the algorithm is stochastic, we ran it 100 times as to reduce the statistical noise. Consequently, we obtained a distribution of Bayesian evidence values and we therefore provide in Tabs.~\ref{tab:resultsDF2}~-~\ref{tab:resultsDF4} and \ref{tab:resultsDF44} the median of the $\mathcal{B}^{\,i}_j$ distribution and the associated error. The Bayes factor can be finally used for the model selection according to the Jeffreys scale \cite{Jeffreys:1939xee}: for $\ln \mathcal{B}^{\,i}_{j}<0$ there is evidence against the model $\mathcal{M}_{i}$, hence the reference model $\mathcal{M}_{j}$ is statistically favored; for $0<\ln\mathcal{B}^{\,i}_{j} <1$ there is no significant evidence in favor of the model $\mathcal{M}_{i}$; for $1<\ln\mathcal{B}^{\,i}_{j} <2.5$ the evidence is substantial; for $2.5< \ln\mathcal{B}^{\,i}_{j} <5$ the evidence is strong; for $\ln\mathcal{B}^{\,i}_{j} >5$ the evidence is decisive.

\section{Results and Discussion}\label{sec:results}

The results of our statistical analysis are reported in Tab.~\ref{tab:resultsDF2} for NGC 1052-DF2, in Tab.~\ref{tab:resultsDF4} for NGC 1052-DF4 and Tab.~\ref{tab:resultsDF44} for Dragonfly 44. The estimates of both the astrophysical and the non-local parameters are provided, together with the Bayes ratio for each of the investigated scenarios. For both the dark matter lacking UDGs, NGC 1052-DF2 and DF4, the reference case for the Bayes ration is the model based on the standard GR with only a stellar component. Concerning the DM dominated UDG, Dragonfly 44, the chosen reference model is characterized by both the stellar and dark matter components, which co-evolve within a General Relativistic framework according to the SHMR relation. Note that in all cases a constant anisotropy profile is always preferred in light of the observational data.

The GR scenarios have therefore been investigated to have a reference model to compare the non-local gravity results with. Moreover, carrying out the standard GR analysis enables us to validate our analysis pipeline by cross-checking the modeling and statistical analysis algorithm against existing literature results.

\subsection{Dark matter lacking UDGs}

Starting from the General Relativity framework, the results for NGC 1052-DF2 are in excellent agreement with those of \cite{Wasserman_2018_df2}, whose analysis adopts a constant anisotropy profile and is carried out both with and without the SHMR relation. Although some differences about the prior choices, the inference yields parameter constraints that are all consistent within the $1\sigma$ confidence level (CL). Also our estimate for the spatially averaged velocity dispersion along the line-of-sight, $\langle \sigma_{\text{los}} \rangle = 6.02^{+1.21}_{-1.19}$ km s$^{-1}$, shows full concordance with that of \cite{Wasserman_2018_df2}. It is worth noticing, however, that our estimate for the anisotropy parameter $\beta_c$ is remarkably lower than that of \cite{Wasserman_2018_df2}, but still the large error bars keep the tension within $1\sigma$. Furthermore, the results from our statistical analysis match the ones obtained in \cite{Laudato:2022vmq}, thus ensuring the validity of our analysis pipeline for NGC 1052-DF2. 

For what concerns NGC 1052-DF4, our findings show good agreement with \cite{van_Dokkum_2019_df4, shen2023confirmation}, whose analyses rule out any dominant dark matter component. The spatially averaged velocity dispersion that we obtain in our fiducial scenario is $\langle \sigma_{\text{los}} \rangle = 5.56^{+1.45}_{-1.32}$ km s$^{-1}$, which is consistent with the updated result of \cite{shen2023confirmation} within the $1\sigma$ uncertainties. Moreover, the constraints provided by our statistical analysis match those of \cite{Laudato:2022drz}, and our algorithm for NGC 1052-DF4 is validated accordingly. 

\begin{figure*}
\centering
\caption{Velocity dispersion profiles of NGC 1052-DF2 in the framework of General Relativity (left panels) and Non-local gravity (right panels). Black dots and bars are the observational data from \cite{van_Dokkum_2018_ml}, with uncertainties. Colored dashed lines and shaded regions respectively are the median and the $1\sigma$ confidence region of the $\sigma_{eff}$ profile, derived from Eq.~(\ref{theoPred}).}\label{fig:profile_plots_DF2}
\includegraphics[width=19cm]{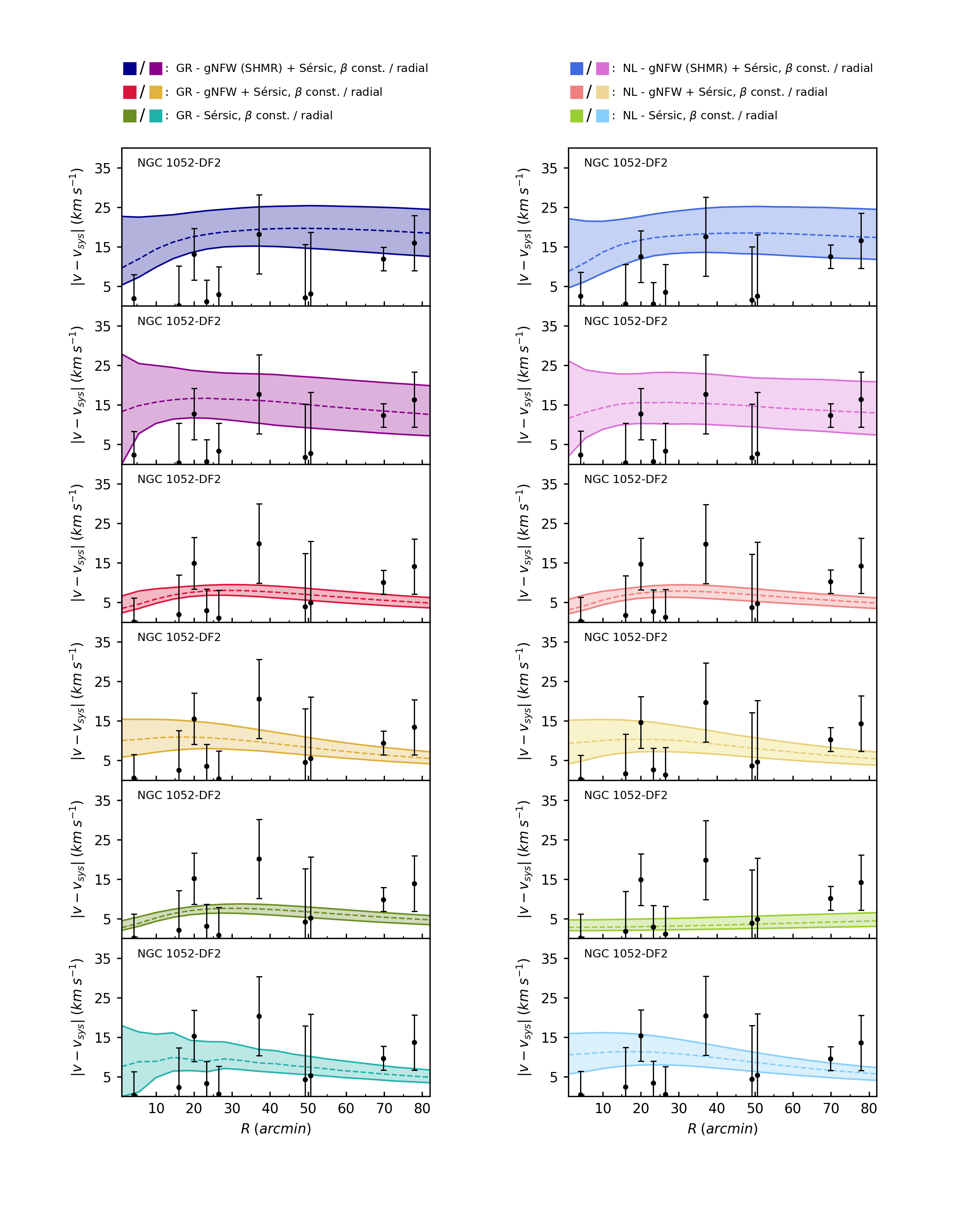}
\end{figure*}

\begin{figure*}
\centering
\caption{Velocity dispersion profiles of NGC 1052-DF4 in the framework of General Relativity (left panels) and Non-local gravity (right panels). Black dots and bars are the observational data from \cite{van_Dokkum_2019_df4}, with uncertainties. Colored dashed lines and shaded regions respectively are the median and the $1\sigma$ confidence region of the $\sigma_{eff}$ profile, derived from Eq.~(\ref{theoPred}).}\label{fig:profile_plots_DF4}
\includegraphics[width=19cm]{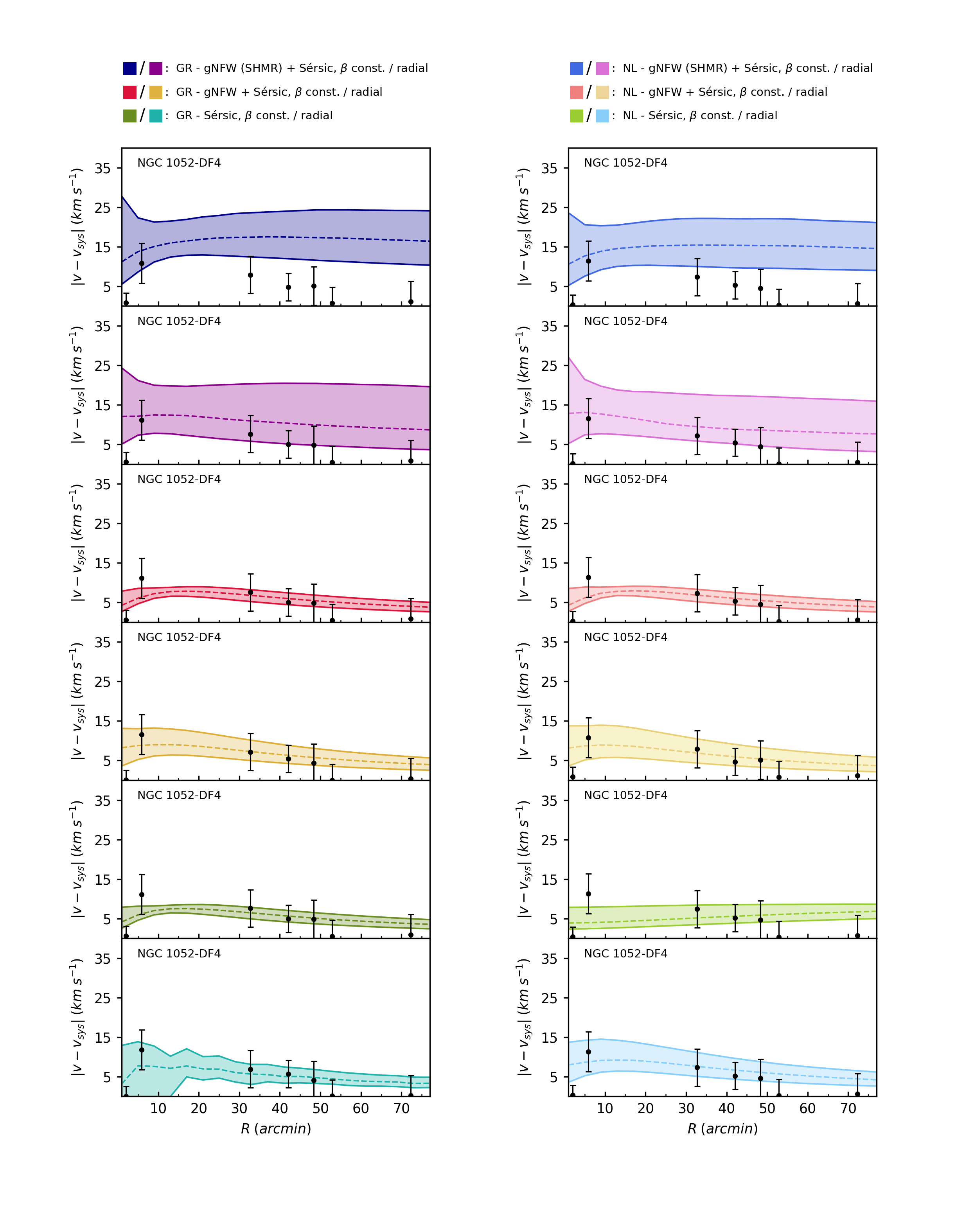}
\end{figure*}

Looking both at the Bayes ratios and the left panels of Fig.~(\ref{fig:profile_plots_DF2}) and Fig.~(\ref{fig:profile_plots_DF4}), it is manifest that the kinematic data of NGC 1052-DF2 and DF4 are in strong tension with the standard galaxy formation paradigm, which predicts the presence of a dominant dark matter component building the halo within which the baryons should accrete. The cases where the SHMR prior is applied are indeed highly disfavoured, since the velocity dispersion is strongly overestimated: $\langle \sigma_{\text{los}} \rangle = 17.73_{-4.79}^{+6.67}$ km s$^{-1}$ and $\langle \sigma_{\text{los}} \rangle = 15.07_{-6.10}^{+7.52}$ km s$^{-1}$, for NGC 1052-DF2; $\langle \sigma_{\text{los}} \rangle = 16.42^{+7.22}_{-5.20}$ km s$^{-1}$ and $\langle \sigma_{\text{los}} \rangle = 10.72^{+9.54}_{-5.10}$ km s$^{-1}$, for NGC 1052-DF4, in the constant and the radial anisotropy scenarios respectively. Moreover, when the gNFW parameters are left free, the best fit estimate of the dark matter content is at least $4.5$ orders of magnitude smaller than expected, hence completely subdominant with respect to baryons. Our analysis also highlights that the velocity dispersions of both UDGs are consistent with the complete absence of dark matter, as can be clearly seen by the results of the Bayesian model selection. This result agrees with the kinematic analysis of \cite{Wasserman_2018_df2, van_Dokkum_2019_df4, shen2023confirmation}. The existence of such dark matter lacking UDGs poses a severe challenge for the standard galaxy formation paradigm, thus stimulating a fruitful discussion within the astrophysiscs community about the possible formation mechanisms \cite{Wright_2021, Jackson_2021:abc, Lee:2021wfv, Trujillo_Gomez_2021, vanDokkum:2022zdd, Ogiya:2022god, Benavides_2023}.

As for the anisotropy profile, the estimates of the anisotropy parameters $\beta_c$, $\beta_0$ and $\beta_\infty$ are all negative. As a consequence, our statistical analysis clearly favours a tangential velocity dispersion, with $\sigma_t$ dominating over the radial component. It is also worth mentioning that the unconstrained scale parameter $r_a$ drifts out to extremely high values in our MCMCs. Since the radial part of the second profile in Eq.~(\ref{anisotropyProfile}) is washed out for large values of $r_a$, our results suggest that the kinematic data of NGC 1052-DF2 and DF4 prefer a fully constant anisotropy profile. This statement is also supported by the Bayes ratios in Tab.~\ref{tab:resultsDF2} and Tab.~\ref{tab:resultsDF4}.

Moving to the results in the non-local gravity framework, it should be noted immediately that the constraints on the whole set of astrophysical parameters remain unaltered. This goes hand in hand with the fact that we are unable to set proper constrains on the non-local length scales, $r_\eta$ and $r_\xi$. Lower bounds are indeed established, representing the minimum values that the non-local parameters can reach without significantly affecting the GR-like fit. In the case of the dark matter lacking UDGs, the lower bounds on the non-local lengths $r_\eta$ and $r_\xi$ show some similarities. In fact, the highest value for the lower bound on $r_\eta$ is obtained in the scenario in which a SHMR relation is applied to the stellar and the dark matter distribution, being  $\log\,r_\eta > -4.94$ for DF2 and $\log\,r_\eta > -4.44$ for DF4. The lowest values, instead, correspond to the case in which we remove the SHMR prior, retaining stellar and dark matter in the energy budget and the radial anisotropy profile, having $\log\,r_\eta > -8.22$ for DF2 and $\log\,r_\eta > -7.64$ for DF4. As far as the lower bounds on $r_\xi$ are concerned, the highest values are observed in the same scenarios as above, with  $\log\,r_\xi > -3.74$ for DF2 and $\log\,r_\xi > -3.08$ for DF4, while the lowest values are $\log\,r_\xi > -8.43$ for DF2 and $\log\,r_\xi > -8.38$) for DF4. Given the $68\%$ CL values reported in Tabs.~\ref{tab:resultsDF2} and \ref{tab:resultsDF4}, the non-local corrections to the Newtonian potential in Eq.~(\ref{phi}) are of the order $\phi_{\text{NL}}$ {\footnotesize$\lesssim$} $0.01 \, \phi_{\text{GR}}$.

Thus, we can conservatively conclude that, even though the recent observations of DM lacking galaxies may pose a threat to those ETGs whose goal is to play the role of the dark matter, UDGs do not represent the deathblow for the alternatives to GR which are claimed by many. The dark energy models based on modifications of gravity can be tested against these anomalous systems, and the observational data can be actually fitted at the same statistical level of GR. The non-local gravity model that has been analysed throughout this work is a clear example of a DE model passing the stress-test posed by the UDG kinematics, as no spoiling effect emerged. Being consistent with zero, the Bayes ratios in Tab.~\ref{tab:resultsDF2} and Tab.~\ref{tab:resultsDF4} indeed show the statistical equivalence of the GR and the non-local fiducial cases, namely the ones with stars only and a constant anisotropy profile. Moreover, despite being statistically disfavoured, the Bayesian model selection for the other cases provides almost equivalent results for the GR scenarios and their non-local counterparts.

\subsection{Dark matter dominated UDG} 

Concerning Dragonfly 44, the results of our analysis match those of \cite{vanDokkum:2019fdc}, where the galaxy is modelled through a less flexible cored-NFW profile, with $\gamma$ almost fixed at $0.3$, and a constant anisotropy profile. It is worth noticing that both the dark matter and the stellar parameters agree within the $1\sigma$ uncertainties, but the stellar ones are poorly constrained and simply follow the priors. As for the anisotropy parameter $\beta_c$, we found a slightly lower value than \cite{vanDokkum:2019fdc}. Although the results are consistent within $1\sigma$, our value deviates from a fully isotropic velocity dispersion towards a tangential anisotropy. Moreover, despite showing the same skewed posterior's shape in the direction of negative values, the constraint on $\beta_c$ from the re-analysis of \cite{Wasserman_2019} lays outside our $68\%$ CL. Neverthless, our results are in perfect agreement with those of \cite{Laudato:2022drz}, which uses the same galaxy modeling, hence our algorithm for Dragonfly 44 is validated against an independent analysis.

\begin{figure*}
\centering
\caption{Velocity dispersion profiles of Dragonfly 44 in the framework of General Relativity (left panels) and Non-local gravity (right panels). Black dots and bars are the observational data from \cite{vanDokkum:2019fdc}, with uncertainties. Colored dashed lines and shaded regions respectively are the median and the $1\sigma$ confidence region of the $\sigma_{eff}$ profile, derived from Eq.~(\ref{theoPred}).}\label{fig:profile_plots_DF44}
\includegraphics[width=19cm]{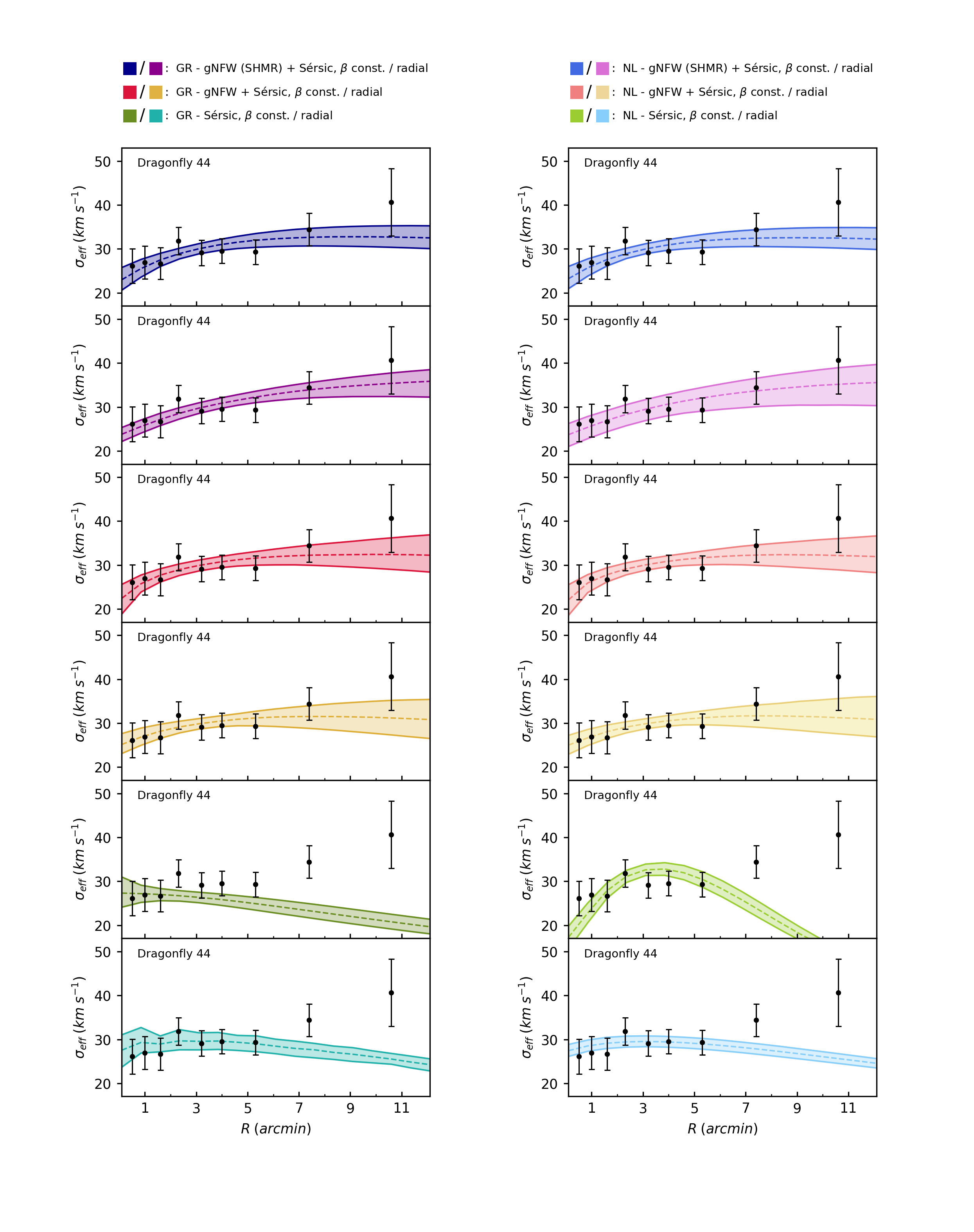}
\end{figure*}

As can be seen from the left panels of Fig.~(\ref{fig:profile_plots_DF44}) and the Bayes ratios of Tab.~\ref{tab:resultsDF44}, in the GR framework the kinematic data of Dragonfly 44 cannot be fitted with stars only, and clearly ask for a dominant dark matter component. When the galaxy is simply modeled through the baryons, the velocity dispersion is strongly underestimated. The spatially averaged value in the constant anisotropy scenario is $\langle \sigma_{\text{los}} \rangle = 24.02^{+1.66}_{-1.63}$ km s$^{-1}$, and the profile shows a descending trend in contrast with the rising dispersion profile that is observed. Moreover, the estimates of the the distance $D$ and the mass-to-light ratio $\Upsilon^*$ remarkably increase, thus being inconsistent with the observed Coma cluster distance \cite{Thomsen:1997vz} and the stellar population synthesis models \cite{vanDokkum:2019fdc}. In the radial case, we instead obtain $\langle \sigma_{\text{los}} \rangle = 27.65^{+1.74}_{-1.78}$ km s$^{-1}$, with a less steep profile approaching a constant shape. The constraints on $D$ and $\Upsilon^*$ also get back to lower values. This is due to the partial degeneracy between the presence of a DM component and how the radial anisotropy affects the kinematics. However, this latter case is also highly disfavoured. 

When we add the dark matter to the modeling of the UDG internal structure, the spatially averaged velocity dispersion significantly increase: $\langle \sigma_{\text{los}} \rangle = 30.75^{+1.98}_{-1.87}$ km s$^{-1}$ and $\langle \sigma_{\text{los}} \rangle = 31.82^{+1.71}_{-1.93}$ km s$^{-1}$, for DM and baryons co-evolving according to the SHMR relation; $\langle \sigma_{\text{los}} \rangle = 30.51^{+2.43}_{-2.32}$ km s$^{-1}$ and $\langle \sigma_{\text{los}} \rangle = 30.20^{+2.48}_{-2.49}$ km s$^{-1}$, for the halo and the stars evolving independently. The results are given for the constant and the radial anisotropy scenario, respectively, each showing an increasing dispersion profile in good agreement with the kinematic data. Focusing on the constant anisotropy case, it is worth mentioning that the application of the SHMR prior enables us to properly constrain the inner slope of the gNFW profile, namely the $\gamma$ parameter. However, though the tighter constraints on both the DM and the anisotropy parameters, the inclusion of the SHMR prior is slightly disfavoured by the Bayesian model selection. This reflects the $2.1\sigma$ tension that we have found between our results, when both the stellar and the DM parameters are left free to vary in the MCMC, and the SHMR relation provided by \cite{Rodr_guez_Puebla_2017}.

As for the anisotropy parameters, when Dragonfly 44 is properly modeled we find analogous results to those of the DM lacking UDGs, although the estimates of $\beta_c$, $\beta_0$ and $\beta_\infty$ are less negative and closer to an isotropic velocity dispersion. Again, the scale parameter $r_a$ remains unconstrained and drifts out to extremely high values in the MCMCs, thus pointing towards a constant anisotropy profile. This result is confirmed by the Bayes ratios in Tab.~\ref{tab:resultsDF44}, with the radial anisotropy cases that are disfavoured with respect to the constant ones. It is worth noticing that the radial case with stars only is instead preferred to its constant counterpart, since an extremely high value of tangential anisotropy can partially mimic the presence of an additional matter component.

Finally, we discuss the results that we obtain for Dragonfly 44 in the non-local gravity framework.

Similarly to the DM lacking UDGs, the constraints on the astrophysical parameters remain unchanged with respect to the ones derived in GR. This reflects the ill-behaved posteriors of $r_\eta$ and $r_\xi$, showing a constant probability distribution from the upper edge of our flat prior $\log \,r_{\eta, \,\xi} \in \mathcal{U}(-10,10)$ to the lower bounds reported in Tab.~(\ref{tab:resultsDF44}), after which the distribution rapidly declines. For the dark matter dominated DF44, the lower bounds on $r_\eta$ correspond to the case of only stellar matter distribution and constant anisotropy, $\log\,r_\eta > -8.07$, while the upper estimations are for a galaxy made up of both stars and dark matter with constant anisotropy and no SHMR, having $\log\,r_\eta > -4.37$. Interestingly, we were able to set a full constraint on $r_\xi$ in the case of stellar distribution and constant anisotropy, $\log\,r_\xi = -8.61^{+0.29}_{-0.29}$, while upper limit corresponds to the scenario of Stars+DM with constant anisotropy and without the SHMR prior, $\log\,r_\xi > -3.98$.

In terms of non-local contributions to the Newtonian potential, these values translate into $\phi_{\text{NL}}$ {\footnotesize$\lesssim$} $0.15 \times \phi_{\text{GR}}$ when the SHMR prior is applied and $\beta=\beta_c$; and $\phi_{\text{NL}}$ {\footnotesize$\lesssim$} $0.005 \times \phi_{\text{GR}}$, when DM and baryons evolve independently and the anisotropy profile is constant. It is worth noticing that when the SHMR relation is added in the UDG internal structure model, the maximum allowed value of the non-local correction to the gravitational potential is remarkably high. 
This is not a surprise, since when the SHMR relation is applied, the mass estimation is increased by $\sim 35\%$. Such an increase is thus reflected by $\phi_{\text{NL}}$ which, as can be seen in Eq.~(\ref{phi}), depends on $M^{2}$.

The only case in which the posteriors of the non-local radii show a different shape is the one described by solely stars and a constant anisotropy profile. Looking at the first row of the non-local gravity section of Tab.~\ref{tab:resultsDF44}, we can see that a tight constraint on $r_\xi$ is achieved, corresponding to a dominant non-local contribution to the Newtonian potential. In this scenario, the non-local correction together with an high tangential anisotropy are able to partially account for the absence of dark matter in the UDG model, yielding significantly better estimates for the Sérsic parameters with respect to the GR case. However, by visual inspection of the dispersion profile in Fig.~(\ref{fig:profile_plots_DF44}), it is evident that the non-local gravity model can mimic the effect of the dark matter only at small and intermediate scales, whereas it completely fails on larger scales. Although the substantial improvement of the Bayes ratio with respect to the GR counterpart, this scenario is still highly disfavoured.

Concerning the other cases, the Bayesian model selection shows a full equivalence between the analyses within the GR framework and those that have been carried out with the non-local model. As a consequence, our non-locally extended model of gravity clearly pass the stress test posed by the kinematics of the DM dominated UDG, Dragonfly 44, showing no spoiling effects at kinematic level. However, the analysis conducted on this galaxy undeniably highlights that this gravity-based DE model cannot also serve as a dark matter model.

\section{Conclusions}\label{sec:conclusions}

In this paper we have carried out an astrophysical test of a dark energy model based upon a non-local extension of General Relativity, Eq.~(\ref{DWaction}). Specifically, we have leveraged the kinematic data of three Ultra-Diffuse Galaxies: NGC 1052 -DF2 and -DF4, which show an almost complete absence of dark matter, and Dragonfly 44, which is completely dark matter dominated. Through the analysis of these anomalous systems, we have been able to investigate very different gravitational regimes, thus performing a true stress test of the non-local gravity model.

To model the velocity dispersion of the three galaxies, we have used the Jeans equation, Eq.~(\ref{JeansEq}), under the assumption of spherical symmetry and no streaming motion. Moreover, we have considered two different functional forms for the dispersion anisotropy, Eq.~(\ref{anisotropyProfile}): a constant profile and a radial one, with the latter that is clearly disfavoured in light of our data analysis. Concerning the internal structure of the UDGs, we have modeled the star content through a Sérsic profile, Eq.~(\ref{Sersic}), while the dark matter component is described by a generalized NFW profile, Eq.~(\ref{gNFW}), which allows for a cored density distribution. The kinematic data of NGC 1052 -DF2 and -DF4 point towards fully baryonic galaxies, and the dark matter component should be absent or completely subdominant. This result is in strong contrast to the SHMR relation as well as the standard galaxy formation paradigm, thus requiring further investigation of the formation mechanisms of the DM lacking UDGs. The kinematic data of Dragonfly 44 instead ask for a dominant dark matter component, even though our results highlight a slight tension between the SHMR relation and the derived estimates of the gNFW parameters, when they are left free to vary in the MCMC.
We should point out, however, that various possible alternative explanations for Dragonfly 44 are in literature. In \cite{Bogdan:2020frt}, for example, it is claimed that the lack of X-ray emission might be explained through the typical stellar-to-halo mass relation established for dwarf galaxies, instead of assuming a ``dark'' galaxy. In \cite{2021MNRAS.502.5921S}, only $20$ globular clusters are found around Dragonfly 44, instead of $80$ of them as it was previously claimed. Accordingly, the total quantity of dark matter in Dragonfly 44 would be that of regular dwarf galaxies of similar stellar mass. On the other hand, in \cite{Wasserman_2019} it was shown how an ultralight scalar field, acting as DM, could provide new opportunities for the explanation of the dynamics of Dragonfly 44.

The main results of this work are evident from the parameter constraints in Tab.~\ref{tab:resultsDF2}~-~\ref{tab:resultsDF4} and \ref{tab:resultsDF44} as well as from the dispersion profiles in Fig.~\ref{fig:profile_plots_DF2}~-~\ref{fig:profile_plots_DF4} and \ref{fig:profile_plots_DF44}. The non-local gravity model does not affect the kinematic predictions for the three galaxies, hence no spoiling effects emerge. Focusing on the fiducial case for each UDG, namely stars only for NGC 1052 -DF2 and -DF4 and baryons and dark matter co-evolving according to the SHMR relation for Dragonfly 44, we can see that the estimates of the astrophysical parameters are not affected by the non-local gravitational corrections. As long as the non-local length scales are larger than the lower bounds reported in the aforementioned Tables, the non-local terms are indeed completely subdominant with respect to the GR contribution to the Newtonian potential.

\begin{figure*}
\centering
\caption{Dependence of the non-local radii on the mass of the astrophysical system. In pink, the lower bound from the analysis of the orbits of the star S2 around the galactic centre \cite{Dialektopoulos:2018iph}; in green, the constraints from this work; in blue, the lower bounds from the analysis of the galaxy cluster lensing of the CLASH sample \cite{Bouche:2022jts}. \vspace{10pt}}\label{fig:nonlocal_radius}
\includegraphics[width=12cm]{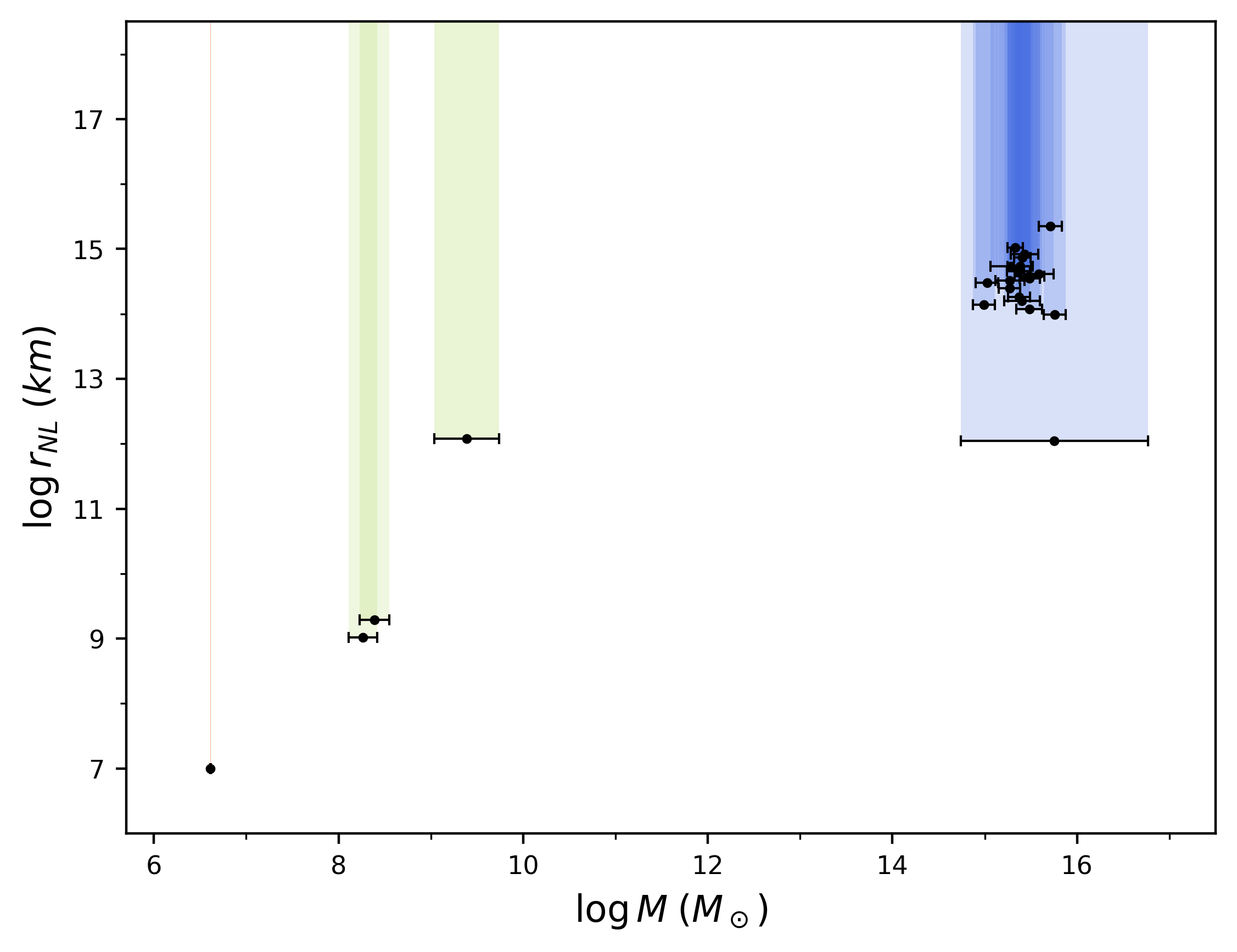}
\end{figure*}

The allowed values for the non-local radii that our analysis yields are shown in Fig.~\ref{fig:nonlocal_radius}, together with the results of previous analyses at galaxy cluster \cite{Bouche:2022jts} and galactic centre scales \cite{Dialektopoulos:2018iph}. The lower bounds derived in this work confirm the correlation between the non-local radii and the mass of the gravitationally bound system under consideration. This result can be interpreted in two ways: on the one hand, the correlation may results from the sensitivity of the data coming from very different scales. Therefore, the constraints from larger structures are consistent with those from smaller astrophysical systems, and the more massive are the structures and the tighter are the derived lower bounds. On the other hand, the correlation might be intrinsic to the nature of the non-local radii. The interaction length scales $r_\eta$ and $r_\xi$ may indeed represent two gravitational radii that the non-local gravity model provides in addition to the Schwarzschild one \cite{Capozziello:2007ec}. In this scenario, $r_\eta(M)$ and $r_\xi(M)$ are two fundamental lengths that govern the kinematics of the cosmic structures.

Finally, we can conclude that the non-local model under consideration brilliantly aced the test posed by the kinematics of Ultra-Diffuse Galaxies as long as it serves as a dark energy model. On the other hand, the analysis of Dragonfly 44 clearly shows that the non-local gravity scenario cannot replace the dark matter, which is still necessary to explain the properties of the gravitationally bound structures of our Universe.

\section*{Acknowledgments}
This article is based upon work from COST Action CA21136 Addressing observational tensions in cosmology with systematic and fundamental physics (CosmoVerse) supported by COST (European Cooperation in Science and Technology). FB, SC and CDS acknowledge the support of {\it Istituto Nazionale di Fisica Nucleare} (INFN), {\it iniziativa specifica} QGSKY .

\bibliographystyle{apsrev4-2}
\bibliography{main}

\newpage
{\renewcommand{\tabcolsep}{2.5mm}
{\renewcommand{\arraystretch}{2.}
\begin{table*}
\begin{minipage}{\textwidth}
\huge
\centering
\caption{Results from the statistical analysis of the dark matter lacking UDG, NGC 1052-DF2. Unconstrained parameters are labelled with $U$. The Bayes factor has been computed using the GR case with the star component solely as reference model (with constant anisotropy profile).}\label{tab:resultsDF2}
\vspace{5pt}
\resizebox*{\textwidth}{!}{
\begin{tabular}{c|c|cc|cccc|ccc|cc|c}
\hline
& Galactic parameters
& \multicolumn{2}{c|}{Sérsic parameters}
& \multicolumn{4}{c|}{Anisotropy parameters}
& \multicolumn{3}{c|}{gNFW parameters} 
& \multicolumn{2}{c|}{Non-local parameters} 
& Bayes factor \\
& $v_{sys}$ & $D$ & $\Upsilon_*$ & $\beta_c$ & $\beta_0$ & $\beta_\infty$ & $r_a$ & $c_{200}$ & $\log M_{200}$ & $\gamma$ & $\log r_\eta$ & $\log r_\xi$ & $\ln\mathcal{B}^{\,i}_j$ \\
& (km s$^{-1}$) & (Mpc) & & & & & (kpc) & & ($M_\odot$) & & (kpc) & (kpc) & \\
\hline
\hline
& \multicolumn{13}{c}{\textbf{General Relativity}} \\
\hline
\hline
\multirow{2}{*}{Stars} & $1804.14^{+2.61}_{-2.57}$ & $22.13^{+1.19}_{-1.15}$ & $1.78^{+0.49}_{-0.46}$ & $-3.81^{+2.67}_{-3.57}$ & $-$ & $-$ & $-$ & $-$ & $-$ & $-$ & $-$ & $-$ & $0$ \\ & $1804.33^{+3.07}_{-3.01}$ & $22.13^{+1.21}_{-1.21}$ & $1.77^{+0.49}_{-0.48}$ & $-$ & $-2.35^{+1.65}_{-2.78}$ & $-1.13^{+1.23}_{-2.42}$ & $U$ & $-$ & $-$ & $-$ & $-$ & $-$ & $-0.50^{+0.04}_{-0.03}$ \\
\hline
\multirow{2}{*}{Stars + DM} & $1803.91^{+2.73}_{-2.55}$ & $22.16^{+1.18}_{-1.26}$ & $1.79^{+0.46}_{-0.46}$ & $-3.46^{+2.39}_{-3.67}$ & $-$ & $-$ & $-$ & $23.53^{+9.06}_{-8.19}$ & $<6.49$ & $U$ & $-$ & $-$ & $-0.30^{+0.03}_{-0.03}$ \\ & $1804.19^{+2.91}_{-3.03}$ & $22.02^{+1.20}_{-1.11}$ & $1.66^{+0.49}_{-0.50}$ & $-$ & $-2.90^{+1.94}_{-3.42}$ & $-1.06^{+1.23}_{-2.64}$ & $U$ & $24.45^{+9.55}_{-8.71}$ & $<5.69$ & $U$ & $-$ & $-$ & $-0.91^{+0.04}_{-0.03}$ \\
\hline
\multirow{2}{*}{Stars + DM (SHMR)} & $1802.09^{+3.75}_{-3.90}$ & $21.88^{+1.15}_{-1.16}$ & $1.55^{+0.53}_{-0.53}$ & $-1.86^{+1.83}_{-3.78}$ & $-$ & $-$ & $-$ & $8.28^{+3.28}_{-2.36}$ & $10.81^{+0.16}_{-0.18}$ & $<0.36$ & $-$ & $-$ & $-2.68^{+0.05}_{-0.03}$ \\ & $1802.86^{+3.83}_{-3.71}$ & $21.97^{+1.23}_{-1.23}$ & $1.61^{+0.51}_{-0.52}$ & $-$ & $-0.23^{+0.36}_{-1.65}$ & $-0.83^{+0.66}_{-0.75}$ & $U$ & $9.70^{+4.40}_{-3.12}$ & $10.83^{+0.15}_{-0.17}$ & $<0.50$ & $-$ & $-$ & $-1.30^{+0.05}_{-0.04}$ \\
\hline
\hline
& \multicolumn{13}{c}{\textbf{Non-local gravity}} \\
\hline
\hline
\multirow{2}{*}{Stars} & $1803.85^{+2.59}_{-2.70}$ & $22.13^{+1.16}_{-1.17}$ & $1.80^{+0.48}_{-0.46}$ & $-3.79^{+2.60}_{-3.63}$ & $-$ & $-$ & $-$ & $-$ & $-$ & $-$ & $>-7.19$ & $>-7.76$ & $-0.03^{+0.03}_{-0.03}$ \\ & $1804.42^{+3.10}_{-2.85}$ & $22.02^{+1.21}_{-1.30}$ & $1.72^{+0.48}_{-0.51}$ & $-$ & $-3.31^{+2.12}_{-3.57}$ & $-1.26^{+1.37}_{-2.96}$ & $U$ & $-$ & $-$ & $-$ & $>-7.68$ & $>-7.83$ & $-0.80^{+0.04}_{-0.04}$ \\
\hline
\multirow{2}{*}{Stars + DM} & $1803.73^{+2.76}_{-2.68}$ & $22.19^{+1.26}_{-1.18}$ & $1.83^{+0.46}_{-0.48}$ & $-4.01^{+2.69}_{-3.30}$ & $-$ & $-$ & $-$ & $24.97^{+8.13}_{-8.81}$ & $<6.17$ & $U$ & $>-8.10$ & $>-7.50$ & $-0.33^{+0.03}_{-0.03}$ \\ & $1804.19^{+3.01}_{-3.16}$ & $22.42^{+1.05}_{-1.20}$ & $1.59^{+0.49}_{-0.46}$ & $-$ & $-3.04^{+2.00}_{-2.98}$ & $-1.16^{+1.27}_{-1.75}$ & $U$ & $25.14^{+8.98}_{-8.27}$ & $<5.72$ & $U$ & $>-8.22$ & $>-8.43$ & $-1.04^{+0.03}_{-0.03}$ \\
\hline
\multirow{2}{*}{Stars + DM (SHMR)} & $1802.21^{+3.57}_{-3.93}$ & $21.91^{+1.21}_{-1.25}$ & $1.58^{+0.48}_{-0.54}$ & $-1.80^{+1.89}_{-3.83}$ & $-$ & $-$ & $-$ & $8.80^{+3.62}_{-2.53}$ & $10.82^{+0.17}_{-0.18}$ & $<0.34$ & $>-5.05$ & $>-3.74$ & $-1.81^{+0.04}_{-0.04}$ \\ & $1802.44^{+3.41}_{-3.66}$ & $22.17^{+1.33}_{-1.22}$ & $1.65^{+0.44}_{-0.56}$ & $-$ & $-0.33^{+0.49}_{-2.19}$ & $-0.91^{+0.80}_{-1.37}$ & $U$ & $9.63^{+3.74}_{-3.09}$ & $10.84^{+0.15}_{-0.17}$ & $<0.45$ & $>-4.94$ & $>-4.10$ & $-0.95^{+0.04}_{-0.05}$ \\
\hline
\end{tabular}}
\end{minipage}
\end{table*}}}

{\renewcommand{\tabcolsep}{2.5mm}
{\renewcommand{\arraystretch}{2.}
\begin{table*}
\begin{minipage}{\textwidth}
\huge
\centering
\caption{Results from the statistical analysis of the dark matter lacking UDG, NGC 1052-DF4. Unconstrained parameters are labelled with $U$. The Bayes factor has been computed using the GR case with the star component solely as reference model (with constant anisotropy profile).}\label{tab:resultsDF4}
\vspace{5pt}
\resizebox*{\textwidth}{!}{
\begin{tabular}{c|c|cc|cccc|ccc|cc|c}
\hline
& Galactic parameters
& \multicolumn{2}{c|}{Sérsic parameters}
& \multicolumn{4}{c|}{Anisotropy parameters}
& \multicolumn{3}{c|}{gNFW parameters} 
& \multicolumn{2}{c|}{Non-local parameters} 
& Bayes factor \\
& $v_{sys}$ & $D$ & $\Upsilon_*$ & $\beta_c$ & $\beta_0$ & $\beta_\infty$ & $r_a$ & $c_{200}$ & $\log M_{200}$ & $\gamma$ & $\log r_\eta$ & $\log r_\xi$ & $\ln\mathcal{B}^{\,i}_j$ \\
& (km s$^{-1}$) & (Mpc) & & & & & (kpc) & & ($M_\odot$) & & (kpc) & (kpc) & \\
\hline
\hline
& \multicolumn{13}{c}{\textbf{General Relativity}} \\
\hline
\hline
\multirow{2}{*}{Stars} & $1446.03^{+2.35}_{-2.44}$ & $22.03^{+1.22}_{-1.20}$ & $1.92^{+0.50}_{-0.47}$ & $-1.64^{+1.57}_{-3.42}$ & $-$ & $-$ & $-$ & $-$ & $-$ & $-$ & $-$ & $-$ & $0$ \\ & $1445.31^{+2.66}_{-2.66}$ & $22.09^{+1.24}_{-1.24}$ & $1.87^{+0.56}_{-0.49}$ & $-$ & $-1.03^{+1.12}_{-2.72}$ & $-0.63^{+0.90}_{-2.08}$ & $U$ & $-$ & $-$ & $-$ & $-$ & $-$ & $-0.24^{+0.04}_{-0.04}$ \\
\hline
\multirow{2}{*}{Stars + DM} & $1445.99^{+2.22}_{-2.32}$ & $22.02^{+1.26}_{-1.19}$ & $1.96^{+0.49}_{-0.51}$ & $-1.89^{+1.76}_{-3.36}$ & $-$ & $-$ & $-$ & $24.63^{+8.86}_{-7.35}$ & $<5.30$ & $U$ & $-$ & $-$ & $-0.32^{+0.04}_{-0.03}$ \\ & $1445.54^{+2.49}_{-2.97}$ & $22.08^{+1.21}_{-1.29}$ & $1.86^{+0.46}_{-0.53}$ & $-$ & $-1.44^{+1.34}_{-3.10}$ & $-0.79^{+1.00}_{-2.15}$ & $U$ & $25.41^{+8.58}_{-8.27}$ & $<5.31$ & $U$ & $-$ & $-$ & $-0.61^{+0.04}_{-0.04}$ \\
\hline
\multirow{2}{*}{Stars + DM (SHMR)} & $1446.26^{+4.15}_{-4.68}$ & $21.79^{+1.23}_{-1.17}$ & $1.91^{+0.50}_{-0.52}$ & $>-1.51$ & $-$ & $-$ & $-$ & $8.96^{+3.56}_{-2.67}$ & $10.79^{+0.16}_{-0.17}$ & $<0.43$ & $-$ & $-$ & $-5.01^{+0.04}_{-0.04}$ \\ & $1445.97^{+3.84}_{-3.71}$ & $21.90^{+1.25}_{-1.20}$ & $1.93^{+0.52}_{-0.54}$ & $-$ & $-0.01^{+0.31}_{-1.03}$ & $-0.70^{+0.64}_{-0.31}$ & $U$ & $9.56^{+3.98}_{-2.97}$ & $10.79^{+0.16}_{-0.17}$ & $<0.58$ & $-$ & $-$ & $-1.95^{+0.05}_{-0.05}$ \\
\hline
\hline
& \multicolumn{13}{c}{\textbf{Non-local gravity}} \\
\hline
\hline
\multirow{2}{*}{Stars} & $1445.83^{+2.23}_{-2.37}$ & $22.17^{+1.16}_{-1.17}$ & $1.92^{+0.49}_{-0.46}$ & $-1.72^{+1.65}_{-3.38}$ & $-$ & $-$ & $-$ & $-$ & $-$ & $-$ & $>-7.46$ & $>-7.63$ & $-0.01^{+0.04}_{-0.03}$ \\ & $1445.76^{+2.68}_{-2.77}$ & $22.17^{+1.11}_{-1.19}$ & $1.88^{+0.52}_{-0.54}$ & $-$ & $-2.06^{+1.73}_{-3.63}$ & $-0.87^{+0.97}_{-2.40}$ & $U$ & $-$ & $-$ & $-$ & $>-7.39$ & $>-8.38$ & $-0.53^{+0.04}_{-0.04}$ \\
\hline
\multirow{2}{*}{Stars + DM} & $1445.71^{+2.28}_{-2.22}$ & $22.14^{+1.16}_{-1.26}$ & $2.00^{+0.49}_{-0.50}$ & $-1.62^{+1.59}_{-3.40}$ & $-$ & $-$ & $-$ & $26.19^{+8.22}_{-8.24}$ & $<5.34$ & $U$ & $>-7.43$ & $>-7.47$ & $-0.27^{+0.04}_{-0.03}$ \\ & $1446.30^{+2.47}_{-2.77}$ & $21.91^{+1.23}_{-1.18}$ & $1.89^{+0.50}_{-0.52}$ & $-$ & $-1.23^{+1.20}_{-3.03}$ & $-0.75^{+0.97}_{-3.03}$ & $U$ & $25.37^{+8.21}_{-8.32}$ & $<5.36$ & $U$ & $>-7.64$ & $>-8.10$ & $-0.48^{+0.04}_{-0.04}$ \\
\hline
\multirow{2}{*}{Stars + DM (SHMR)} & $1445.71^{+4.42}_{-4.29}$ & $21.95^{+1.30}_{-1.19}$ & $1.88^{+0.51}_{-0.51}$ & $>-1.34$ & $-$ & $-$ & $-$ & $8.89^{+3.31}_{-2.67}$ & $10.78^{+0.15}_{-0.17}$ & $<0.47$ & $>-4.97$ & $>-3.08$ & $-3.69^{+0.06}_{-0.05}$ \\ & $1445.56^{+3.26}_{-3.19}$ & $21.95^{+1.36}_{-1.25}$ & $1.91^{+0.53}_{-0.53}$ & $-$ & $0.06^{+0.33}_{-0.71}$ & $-0.55^{+0.62}_{-0.41}$ & $U$ & $9.63^{+4.06}_{-3.05}$ & $10.78^{+0.15}_{-0.16}$ & $<0.60$ & $>-4.44$ & $>-4.78$ & $-2.04^{+0.05}_{-0.04}$ \\
\hline
\end{tabular}}
\end{minipage}
\end{table*}}}

{\renewcommand{\tabcolsep}{2.5mm}
{\renewcommand{\arraystretch}{2.}
\begin{table*}
\begin{minipage}{\textwidth}
\huge
\centering
\caption{Results from the statistical analysis of the dark matter dominated UDG, Dragonfly 44. Unconstrained parameters are labelled with $U$. The Bayes factor has been computed using the GR case with both the star and the dark matter component as reference model (with constant anisotropy profile, and SHMR relation).}\label{tab:resultsDF44} 
\vspace{5pt}
\resizebox*{\textwidth}{!}{
\begin{tabular}{c|cc|cccc|ccc|cc|c}
\hline
& \multicolumn{2}{c|}{Sérsic parameters} 
& \multicolumn{4}{c|}{Anisotropy parameters}
& \multicolumn{3}{c|}{gNFW parameters} 
& \multicolumn{2}{c|}{Non-local parameters} 
& Bayes factor \\
& $D$ & $\Upsilon_*$ & $\beta_c$ & $\beta_0$ & $\beta_\infty$ & $r_a$ & $c_{200}$ & $\log M_{200}$ & $\gamma$ & $\log r_\eta$ & $\log r_\xi$ & $\ln\mathcal{B}^{\,i}_j$ \\
& (Mpc) & & & & & (kpc) & & ($M_\odot$) & & (kpc) & (kpc) & \\
\hline
\hline
& \multicolumn{12}{c}{\textbf{General Relativity}} \\
\hline
\hline
\multirow{2}{*}{Stars} & $148.84^{+11.90}_{-11.72}$ & $10.34^{+1.38}_{-1.27}$ & $0.10^{+0.13}_{-0.16}$ & $-$ & $-$ & $-$ & $-$ & $-$ & $-$ & $-$ & $-$ & $-49.61^{+0.03}_{-0.03}$ \\ & $112.69^{+12.47}_{-11.79}$ & $2.10^{+0.36}_{-0.31}$ & $-$ & $<-8.10$ & $-0.29^{+0.74}_{-1.11}$ & $U$ & $-$ & $-$ & $-$ & $-$ & $-$ & $-4.54^{+0.04}_{-0.04}$ \\
\hline
\multirow{2}{*}{Stars + DM} & $101.54^{+14.24}_{-14.48}$ & $1.63^{+0.39}_{-0.33}$ & $-0.50^{+0.49}_{-3.26}$ & $-$ & $-$ & $-$ & $10.21^{+5.12}_{-4.06}$ & $10.80^{+2.33}_{-0.73}$ & $U$ & $-$ & $-$ & $0.27^{+0.04}_{-0.04}$ \\ & $98.03^{+14.50}_{-15.18}$ & $1.67^{+0.52}_{-0.36}$ & $-$ & $>-3.76$ & $-1.00^{+1.10}_{-2.07}$ & $U$ & $11.05^{+4.13}_{-3.57}$ & $10.23^{+0.64}_{-0.63}$ & $<0.67$ & $-$ & $-$ & $-0.64^{+0.04}_{-0.04}$ \\
\hline
\multirow{2}{*}{Stars + DM (SHMR)} & $100.18^{+13.76}_{-13.63}$ & $1.55^{+0.40}_{-0.32}$ & $-0.36^{+0.27}_{-0.42}$ & $-$ & $-$ & $-$ & $10.75^{+3.13}_{-2.71}$ & $10.98^{+0.14}_{-0.15}$ & $<0.69$ & $-$ & $-$ & $0$ \\ & $96.54^{+15.11}_{-15.71}$ & $1.55^{+0.42}_{-0.32}$ & $-$ & $-1.39^{+1.12}_{-2.72}$ & $-0.55^{+0.67}_{-1.97}$ & $U$ & $8.46^{+3.34}_{-2.27}$ & $10.97^{+0.16}_{-0.18}$ & $<0.47$ & $-$ & $-$ & $-0.60^{+0.04}_{-0.04}$ \\
\hline
\hline
& \multicolumn{12}{c}{\textbf{Non-local gravity}} \\
\hline
\hline
\multirow{2}{*}{Stars} & $104.20^{+13.76}_{-13.68}$ & $1.72^{+0.47}_{-0.35}$ & $-2.41^{+0.82}_{-1.36}$ & $-$ & $-$ & $-$ & $-$ & $-$ & $-$ & $>-8.07$ & $-8.61^{+0.29}_{-0.29}$ & $-11.90^{+0.04}_{-0.03}$ \\ & $113.17^{+12.94}_{-13.11}$ & $2.13^{+0.42}_{-0.35}$ & $-$ & $<-7.85$ & $-0.10^{+0.62}_{-1.11}$ & $U$ & $-$ & $-$ & $-$ & $>-6.65$ & $>-6.79$ & $-4.59^{+0.03}_{-0.03}$ \\
\hline
\multirow{2}{*}{Stars + DM} & $103.37^{+13.99}_{-13.82}$ & $1.53^{+0.39}_{-0.32}$ & $-0.89^{+0.83}_{-3.67}$ & $-$ & $-$ & $-$ & $11.03^{+4.53}_{-3.69}$ & $10.50^{+1.87}_{-0.50}$ & $U$ & $>-4.37$ & $>-4.52$ & $0.25^{+0.04}_{-0.04}$ \\ & $99.73^{+15.56}_{-11.46}$ & $1.58^{+0.48}_{-0.29}$ & $-$ & $>-3.91$ & $-1.14^{+1.16}_{-3.12}$ & $U$ & $11.14^{+4.68}_{-3.64}$ & $10.27^{+0.75}_{-0.54}$ & $<0.56$ & $>-5.13$ & $>-3.98$ & $-0.40^{+0.04}_{-0.04}$ \\
\hline
\multirow{2}{*}{Stars + DM (SHMR)} & $99.99^{+14.15}_{-14.88}$ & $1.55^{+0.38}_{-0.29}$ & $-0.38^{+0.27}_{-0.42}$ & $-$ & $-$ & $-$ & $10.53^{+3.34}_{-2.76}$ & $10.97^{+0.15}_{-0.16}$ & $<0.71$ & $>-4.40$ & $>-5.11$ & $-0.05^{+0.04}_{-0.04}$ \\ & $96.74^{+14.07}_{-13.81}$ & $1.53^{+0.47}_{-0.32}$ & $-$ & $-1.70^{+1.41}_{-3.04}$ & $-0.56^{+0.70}_{-2.76}$ & $U$ & $8.76^{+2.11}_{-2.29}$ & $11.00^{+0.14}_{-0.19}$ & $<0.42$ & $>-4.53$ & $>-4.96$ & $-0.56^{+0.04}_{-0.04}$ \\
\hline
\end{tabular}}
\end{minipage}
\end{table*}}}

\newpage

\newpage

\newpage

\end{document}